\documentclass[aps,prc,preprint,tightenlines,showpacs,preprintnumbers,amsmath,amssymb,superscriptaddress]{revtex4}

\usepackage{psfig}
\usepackage{graphicx}
\usepackage{dcolumn}
\usepackage{bm}

\newcommand{\sla}{ {\hspace{-0.19cm} \setminus} }
\newcommand{\slas}{ {\hspace{-0.19cm} \setminus}\hspace{-0.0cm} }

\newcommand{\be}{\begin{eqnarray}}
\newcommand{\ee}{\end{eqnarray}}

\newcommand{\bc}{\begin{center}}
\newcommand{\ec}{\end{center}}

\newcommand{\npi}{\mbox{$\pi N$}}

\newcommand{\nrho}{\mbox{$\rho N$}}
\newcommand{\neta}{\mbox{$\eta N$}}
\newcommand{\dpi}{\mbox{$\pi \Delta$}}

\begin{document}

\title{Pion-nucleon scattering in a meson-exchange model}

\author{A.M. Gasparyan}
\affiliation{Institut f\"ur Kernphysik (Theorie), Forschungszentrum J\"ulich GmbH,
D-52425 J\"ulich, Germany}
\affiliation{Institute of Theoretical and Experimental Physics, 
117259, B. Cheremushkinskaya 25, Moscow, Russia}
\author{J. Haidenbauer}
\author{C. Hanhart}
\author{J. Speth}
\affiliation{Institut f\"ur Kernphysik (Theorie), Forschungszentrum J\"ulich GmbH,
D-52425 J\"ulich, Germany}

\begin{abstract}
The $\pi N$ interaction is studied within a meson-exchange model  
and in a coupled-channels approach
which includes the channels $\pi N$, $\eta N$, as well as three 
effective $\pi\pi N$ channels namely $\rho N$, $\pi\Delta$, and 
$\sigma N$. Starting out from an earlier model of the J\"ulich 
group systematic improvements in the dynamics and in some 
technical aspects are introduced. 
With the new model an excellent quantitative reproduction 
of the $\pi N$ phase shifts and inelasticity parameters 
in the energy region up to 1.9 GeV and for total angular momenta 
$J\leq3/2$ is achieved. 
Simultaneously, good agreement with data for the 
total and differential $\pi N\to \eta  N$ transition cross sections 
is obtained. The connection of the $\pi N$ dynamics in the $S_{11}$
partial wave with the reaction $\pi N\to \eta  N$ is discussed.
\end{abstract}


\maketitle

%----------------------
 \section{Introduction}
%----------------------

The $\pi N$ interaction is interesting for several reasons. First, it is
one of the main sources of information about the baryon spectrum. Thereby
it serves as a doorway to the understanding of QCD in the non-perturbative
regime - and 
especially of the confining mechanism, which is most important for
binding a system of quarks into a hadron.  For example, experimental information
about the mass, width and decay of baryon resonances serves as a testing
ground for several models of the internal structure of the nucleon and
its excited states. Most of this information is extracted from partial
wave analyses of $\pi N$ scattering data \cite{Koch85,SM95,SP98}.

The $\pi N$ interaction is also interesting by itself. The wealth of
accurate data and the richness of structures shown by them provide
an excellent but also challenging testing ground for any model 
description in terms of effective degrees of freedom, e.g., for
chiral perturbation theory \cite{Mojzis98,Fettes01} but also for the more 
conventional meson-exchange picture \cite{Pearce91,Gross93,Krehl00}. 

Finally, the $\pi N$ interaction is an important ingredient in many other
hadronic reactions and in particular for the meson production in 
nucleon-nucleon ($NN$) collisions \cite{Machner99,Moskal02}. 
$\pi N$ rescattering is an essential mechanism in the reaction $NN\to NN\pi$ near 
threshold \cite{Hernandez95,Hanhart95,Hanhart98}. 
There are also strong indications that rescattering involving the 
$\pi N$ system plays an important if not dominant role in the
production of the $\eta$ \cite{Batinic97,Wilkin01,Nakayama02,Baru03} 
and $\omega$ mesons \cite{Nakayama98,Nakayama03} 
and even for the associated strangeness production 
($NN\to N\Lambda K$, $NN\to N\Sigma K$) \cite{Gasparian00,Shyam01}. 
Thus, model investigations of such production reactions require solid
information about the corresponding elementary reactions like 
$\pi N \to \eta N$, $\pi N \to \omega N$,
$\pi N \to K \Lambda$, $\pi N \to K \Sigma$, etc. 

Over the last few years, in a series of papers, the J\"ulich group has investigated 
the $\pi N$ interaction in the meson-exchange framework 
\cite{Schuetz94,Schuetz95,Schuetz98,Krehl00}. One of the main novelties of
the model was treating the $\sigma$ and $\rho$-meson $t$-channel exchanges
as correlated two-pion exchange, using the dispersion relations technique. 
The J\"ulich model was originally constructed to describe elastic $\pi N$ data
not far from threshold \cite{Schuetz94}. Later the model was extended to higher
energies by including several inelastic channels namely
three effective $\pi \pi N$ channels ($\sigma N$, $\rho N$, and $\Delta\pi$) and the
$\eta N$ channel \cite{Schuetz98,Krehl00}. The treatment
of correlated $\pi\pi$-exchange was made more
consistent and transparent in Ref. \cite{Krehl99}.
The possibility of generating resonances dynamically was also
systematically studied. It turned out, that only one of them, namely
the Roper resonance ($P_{11}(1440)$), can be understood in this way in
the framework of the J\"ulich $\pi N$ model \cite{Schuetz98,Krehl00}.
Other resonances like $S_{11}(1535)$, $S_{11}(1650)$, $D_{13}(1520)$,
and $\Delta(1232)$ had to be included explicitly.
The latest model provided a good qualitative and in many partial
waves even a quantitative description of $\pi N$ scattering in
the energy region from theshold up to 1.9 Gev. \cite{Krehl00}.

Unfortunately, a further improvement of this model  
by simply introducing further resonances and by including additional 
inelastic channels proved to be impossible due to several reasons.
First of all, in some partial waves the deviation of the model
predictions from the data at higher energies are seemingly not only
due to missing resonance contributions. Already 
the basic (non-resonant or background) contributions
of the model by Krehl et al. \cite{Krehl00} are incompatible 
with the general trend exhibited by the experimental phase shifts.

The second problem is a strong influence of the
$N^*(1650)$ resonance on the low energy $S_{11}$ phase shift.
In fact, it gives the main contribution to this partial wave even at threshold 
-- which is, of course, unphysical. This means, in turn, that any additional
channels that couple to the $N^*(1650)$ resonance will likewise have
a strong influence on the $S_{11}$ phase shift close to threshold,
a certainly undesirable feature. 

Finally, the existing $\pi N$ model yields only an unsatisfactory
description of the inelasticity parameter in the $S_{11}$ partial wave 
and at the same time it overestimates the $\pi N\to \eta N$ transition cross 
section close to the $\eta N$ threshold. These two related problems
are believed to be due to shortcomings in the
treatment of the $\pi\pi N$ channel.

In this context, let us mention 
that the $S_{11}$ partial wave is of particular
importance  for the $\eta N$ and $K \Lambda$ channels close to 
their thresholds.
For $\pi N \to \eta N$ as well as $\pi N \to K \Lambda$
experimental information on the transition cross sections and also
differential cross sections and  polarization observables are
available. An analysis of those data within our
model requires a satisfactory description of the $S_{11}$ $\pi N$
partial wave in the relevant energy range.
Moreover, an adequate description of the $S_{11}$ inelasticity and of
the $\pi N \to \eta N$ transition amplitude is also needed if one 
wishes to investigate $\eta$ production in $NN$ collisions \cite{Baru03}. 
Similarly, 
the $\pi N\to K\Lambda(\Sigma)$ transition amplitude plays an
important role in studies of $\Lambda(\Sigma)$ production in
$NN$ collisions. It is the main ingredient in the production amplitude 
based on the pion rescattering mechanism \cite{Gasparian00}. 
 
In the present work we want to remedy the above-mentioned deficiencies
of the J\"ulich $\pi N$ model \cite{Krehl00}. Thereby we aim at a quantitative
description of the $\pi N$ phase shifts and inelasticities for all
partial waves with $J\leq 3/2$, from threshold up to around 1.9 GeV.
A further and equally important goal is the consistent description of the 
experimental information on the $\pi N \to \eta N$ transition. 
 
The paper is structured in the following way: 
In Sec. II the main ingredients of our $\pi N$ model are described  
with special emphasis on those parts of the dynamics where 
changes and improvements were made.  
For the time being, apart from the $\pi\pi N$ channel
(described effectively via the $\sigma N$, $\rho N$, and $\Delta\pi$ channels)
only the $\eta N$ channel is taken into account. However,
the inclusion of the $K \Lambda$ channel (and even $\omega N$ and $K \Sigma$)
is expected to be straightforward within the new improved model.
In Sec. III we present results for the $\pi N$ elastic
scattering. Specifically, we compare the $\pi N$ phase shifts
and inelasticities of the new model with experimental values and
with the description achieved within the model of Krehl et al. \cite{Krehl00}. 
In addition, and as the main result of our paper 
we examine in detail the transition reaction $\pi N\to \eta N$.
Calculations for the total transition cross section
but also for differential observables are presented. 
Furthermore, we shed some light on peculiar structures which occur
in the $\pi N\to \eta N$ total cross section of our old model, but
also in other models in the literature \cite{Gridnev99,Penner02a,Shklyar03}. 
The paper ends with a short summary.

%----------------------
 \section{Description of the model}
%----------------------
The general framework as well as all technical aspects of the
J\"ulich $\pi N$ model have been thoroughly described in earlier
papers \cite{Schuetz94,Krehl99,Krehl00}. Therefore, we refrain 
from repeating all the details here. Rather we want to give
a brief account of its main features with specific emphasis on 
the new and improved ingredients of the present model. 

Our model of the $\pi N$ interaction is derived within
the meson-exchange framework in time-ordered perturbation theory (TOPT). 
Within the envisaged range of validity of our model of up
to around 1.9 GeV inelasticities play an increasingly important
role, as is evidenced by the results of phase-shift analyses. 
Hence, coupling to reaction channels that are responsible
for those inelasticities have to be taken into account. 
The decay modes of the
nucleon resonances in the energy range under consideration show that
the dominant decay (besides $\pi N$ and $\eta N$ for the $N^*(1535)$)
is the $\pi\pi N$ channel \cite{Hagiwara02}. Since a three-body calculation is much
too complicated for realistic potentials, we represent the $\pi\pi N$
channel by effective two-body channels. 
 In doing this we are guided by
studying strong interactions between two-body clusters of the three-body $\pi\pi
N$ state in the spirit of the formalism of Ref. \cite{Aaron68}.
The dominant clusters are the $\Delta$ in the $\pi N$
interaction, the $\rho$ in the vector-isovector $\pi\pi$ interaction and the
strong correlation in the scalar-isoscalar $\pi\pi$ interaction, which we
call $\sigma$. Therefore -- besides the $\pi N$ and $\eta N$ channels
-- we include in our model the reaction channels $\pi\Delta$, $\sigma N$ and $\rho N$.

Accordingly, we have to solve the coupled-channel scattering
equation \cite{Mueller-Groeling90}
\begin{eqnarray}\label{ccstreugl}
\langle \vec{k}' \lambda_3\lambda_4|T^{I}_{\mu\nu}|\vec{k} \lambda_1\lambda_2\rangle
=
&
\langle \vec{k}' \lambda_3\lambda_4|V^{I}_{\mu\nu}|\vec{k} \lambda_1\lambda_2\rangle& +
 \nonumber \\
\sum_{\gamma}\sum_{\lambda'_1,\lambda'_2}\int d^3q
&
\langle \vec{k}' \lambda_3\lambda_4|V^{I}_{\mu\gamma}|\vec{q} \lambda'_1\lambda'_2\rangle& 
\frac{1}{E-W_{\gamma}(q)+i\epsilon} \ 
\langle \vec{q} \lambda'_1\lambda'_2|T^{I}_{\gamma\nu}|\vec{k} \lambda_1\lambda_2\rangle,
\end{eqnarray}
where $\lambda_i,\lambda_{i+2},\lambda'_i, (i=1,2)$ are the helicities of the
baryon and meson in the initial, final and intermediate state, $I$ is
the total isospin of the two body system and $\mu,\nu,\gamma$ are indices 
that label different reaction channels.
$W_{\gamma}(q)=\sqrt{q^2+M_{\gamma}}+\sqrt{q^2+m_{\gamma}}$ where
$m_{\gamma}(M_{\gamma})$ is the mass of the meson (baryon) in the
channel $\gamma$, respectively. We work in the center-of-momentum (cm)
frame and
$k(k')$ are the momenta of the initial (final) baryon, respectively.

The pseudopotential $V^{I}_{\mu\nu}$ (i.e., the interaction between baryon and
meson) that is iterated in Eq. (\ref{ccstreugl}) is constructed
from an effective Lagrangian. Our interaction Lagrangian (see Table
\ref{tablag}) is based on that of Wess and Zumino \cite{Wess67},
which we have supplemented with additional terms for including the
$\Delta$ isobar, the $\omega$, $\eta$, $a_0$ meson and the $\sigma$.
We also have included terms that characterize the coupling of
the resonances $N^*(1535)$, $N^*(1520)$ and $N^*(1650)$ to various
reaction channels. The diagrams that built up the interaction in the 
$\pi N \to \pi N$,
$\pi N \to \eta N$, and $\eta N \to \eta N$ channels 
are shown in Figs. \ref{pingraf} and \ref{etangraf} as an example and
also to introduce our notation. The full set of diagrams, including also
the transitions and interactions in the other reaction channels
($\rho N$, $\sigma N$, $\pi \Delta$), can be found in Ref.~\cite{Krehl00}. 
In that paper one can also find explicit expressions for all the matrix elements 
$\langle \vec{k}' \lambda_3\lambda_4|V^{I}_{\mu\nu}|\vec{k} \lambda_1\lambda_2\rangle$. 

As already indicated in the Introduction, there are some
modifications and improvements in the present model and we want to
summarize them shortly here. First, we now use derivative coupling
for the $S_{11}$ $N^*$ resonances, as demanded by chiral symmetry. 
The corresponding Lagrangians for
the $N^*(S_{11})N\pi$ and $N^*(S_{11})N\eta$ vertices can be found in
table \ref{tablag}. Secondly, we introduce a coupling of the 
$S_{11}$ $N^*$ (1535) resonance to the $\pi \Delta$ channel. Also
this Lagrangian is given in table \ref{tablag}. Finally, the subtraction
constant that appears in the dispersion relations which constitute the
contribution of the correlated $\pi\pi$ exchange in the scalar-isoscalar
($\sigma$) channel \cite{Schuetz94} is not set to zero as in our previous 
models \cite{Schuetz98,Krehl00}, but allowed to assume a finite value. 
Interpreted in terms of effective exchanges this contact term corresponds
to the exchange of a sigma meson with scalar coupling in addition to
the derivative coupling as it occurs now for the sigma exchange stemming
from the subtracted dispersion integral. Note, 
when interpreting the low energy constants $c_i$, as they occur in the chiral
perturbation theory analysis of $\pi N$ scattering, phenomenologically in
terms of resonance exchanges, also both coupling structures of a scalar
to pions can be identified \cite{BKM}.
Explicit expressions for those matrix elements
$\langle \vec{k}' \lambda_3\lambda_4|V^{I}_{\mu\nu}|\vec{k} \lambda_1\lambda_2\rangle$
which differ from the ones employed in our old model~\cite{Krehl00}
can be found in the Appendix.

Mesons and baryons are not point-like particles, but have a finite size.
Therefore the interaction vertices $mmm$ and $mBB$ ($m$=meson,
$B$=Baryon) also have a finite structure which, in our model, is taken into account
by means of form factors. Those form factors are parameterized by the
following analytical forms, in which $\vec{q}$ is the three momentum transfer
carried by the exchanged particle:
\begin{itemize}
\item For meson and baryon exchange
\begin{equation}\label{ffaus}
F(q)=\left(\frac{\Lambda^2-m_x^2}{\Lambda^2+\vec{q}\,^2}\right)^n .
\end{equation}
We use monopole form factors ($n=1$) except for the $\Delta$
exchange, for which the convergence of the integral in Eq. 
(\ref{ccstreugl}) requires a dipole form factor ($n=2$).
\item For the nucleon exchange at the $\pi NN$ vertex
\begin{equation}\label{ffnu}
F(q)=\frac{\Lambda^2-m_N^2}{\Lambda^2-((m_N^2-m_{\pi}^2)/m_N)^2+\vec{q}\,^2}.
\end{equation}
This choice ensures that the nucleon pole and nucleon exchange
contribution cancel each other at the Cheng-Dashen point, which is needed for a 
calculation of the $\Sigma$ term \cite{Schuetz94}.
\item For $N$, $N^*$ and $\Delta$ Pole diagrams
\begin{equation}\label{ffpol}
F(q)=\left(\frac{\Lambda^4+m_R^4}{\Lambda^4+(E_{\gamma}(q)+\omega_{\gamma}(q))^4}\right)^n,
\end{equation}
where $n=1$ is used for $S$ and $P$-wave resonances, and $n=2$
for resonances in higer partial waves.
\item The correlated $\pi\pi$ exchange is supplemented by the form factor
\be\label{sigmaform}
F(p_2,p_4)=\frac{\Lambda^2}{\Lambda^2+\vec{p}_2\,^2}\cdot\frac{\Lambda^2}
{\Lambda^2+\vec{p}_4\,^2}
\ee
Note that this choice differs from the form employed in our previous
$\pi N$ models, where the form factor 
appeared inside the $t'$ integration, cf. Ref. \cite{Schuetz94}.
The particular form we apply in the present work has the following advantages:
i) it does not depend on energy;
ii) it does not modify strongly
the on-shell potential (which is assumed to be fully determined by
the dispersion integrals) as long as the energy is not too high;
iii) it does not change the $t$ dependence of the potential.
\item For the contact interaction in the Wess-Zumino Lagrangian \cite{Wess67}
\begin{equation}\label{ffct}
F(p_2,p_4)=\left(\frac{\Lambda^2+m_4^2}{\Lambda^2+\vec{p}_4\,^2}\frac{\Lambda^2+m_2^2}{\Lambda^2+\vec{p}_2\,^2}\right)^2.
\end{equation}
\end{itemize}

Finally, we want to emphasize that 
the $\Delta$ isobar in the $\pi \Delta$ channel and the $\sigma$ and 
$\rho$ mesons in the $\sigma N$ and $\rho N$ 
channels are not treated as stable particles.  Rather, as already mentioned
above, the $\Delta$, $\sigma$ and $\rho$ 
here stand for $\pi N$ and $\pi \pi$ subsystems with
the quantum numbers of the $P_{33}$ partial wave in the $\pi N$ system and the
$I=J=0$ and $I=J=1$ partial waves in the $\pi \pi$ system, respectively. 
In order to
simulate these, a simplified model for the $P_{33}$ $\pi N$ partial
wave as well as for the $\delta_{00}$ and $\delta_{11}$ 
$\pi \pi$ partial waves was adopted in which pole diagrams, in the framework 
of time-ordered perturbation theory, are iterated
\cite{Schuetz98,Krehl00}.
These models are then used to construct the self-energies of the
$\Delta$, $\sigma$ and $\rho$ which
appear in the propagators of the $\pi \Delta$ and $\sigma N$ intermediate
states in our scattering equation, i.e. we replace 
the two-particle intermediate state propagator for $\pi\Delta$,
$\sigma N$ and $\rho N$ by
\begin{equation}\label{modprop}
\frac{1}{E-W_{\gamma}(q)}\to\frac{1}{E-W_{\gamma}(q)-\Sigma_{\gamma}(E_{sub})},
\end{equation}
where
\begin{eqnarray}
E_{sub}&=&E-\omega_{\pi}(q)-(\sqrt{(M_{\Delta}^o)^2+q^2}-M_{\Delta}^o) \mbox{
for the } \Delta, \nonumber \\
E_{sub}&=&E-E_{N}(q)-(\sqrt{(m_r^o)^2+q^2}-m_r^o)  \mbox{ for } r=\rho,\sigma
\end{eqnarray}
is the energy of the decaying cluster at rest \cite{Schuetz98}. 
The bare masses $M^o_{\Delta}$ and $m_r^o$ are determined by fitting the models 
to the relevant phase shifts of the $\pi N$ and $\pi\pi$ systems, 
cf. Refs. \cite{Schuetz98} and \cite{Krehl00} for details. 
By taking into account the self-energy contibutions we preserve the correct
threshold behavior for the description of pion production in the $\pi N$
system.

The scattering equation (\ref{ccstreugl}) is reduced 
to a set of one-dimensional intergral equations by means of the usual partial wave 
decomposition \cite{Erkelenz74} and then solved numerically by standard 
contour-deformation methods \cite{Aaron66,Cahill71}. 

%----------------------
 \section{Results}
%----------------------

In this section we present the
results of our model for $\pi N$ elastic scattering and 
for the $\pi N\to \pi \eta$
transition in the energy range from $\pi N $
threshold up to 1.9 GeV.
First we discuss the parameters that enter into our model
calculation. Then we present the results for the 
$\pi N$ phase shifts and inelasticities. In particular,
the role of the background and of the
resonance contributions is analyzed. We also
compare the results with those of the
previous version of the model.
Furthermore, we analyse in detail the results
for the $\pi N\to \eta N$ total cross
section and angular distributions and discuss
the role of the background in this 
process.

\subsection{Parameters of the model}

Our model is based on the effective potential,
which was described in section II.

The masses of all the particles appearing in
the model are collected in table \ref{masses}. 
Here one should pay attention to the mass of
the $\sigma$ meson. While the $\sigma$ exchange 
in the $\pi N\to \pi N$ potential is evaluated 
using a dispersion relation, we have another
$t$-channel $\sigma$ exchange in the $\sigma N\to\sigma N$ potential.
In this case we choose the value $m_\sigma=650$ MeV, which
was extracted from a Breit-Wigner parameterization
of the correlated $\pi\pi$-exchange in Ref.~\cite{Durso80} .

Table \ref{bg_param} contains coupling
constants and cutoff parameters of the form factors
for the vertices entering the $t$ and $u$-exchange diagrams
and the contact terms, i.e. those, which constitute 
the background.

Most of the coupling constants have been taken
from other sources. The coupling constants of the pole diagrams are
constrained by values determined from their decay widths, for which we
take the estimates of Ref. \cite{PDG98}.
The parameters which are not fixed from other sources are
shown in boldface. These are the purely phenomenological coupling 
constant at the triple $\sigma$ vertex, $g_{\sigma\sigma\sigma}$, and the subtraction 
constant $A_0$ for the dispersion relation in the $\sigma$ channel.
In addition the cutoff masses are treated as free parameters.
Those free parameters are determined by a fit to the $\pi N$ phase shifts
and inelasticities for $J\leq 3/2$ and the $\pi N\to \eta N$ cross section
in the energy range from threshold to about 1.9 GeV. 
Here we should emphasize that we restrict ourselves to
values of the cutoff masses of about 1-1.5 GeV 
(in some cases up to 2 GeV for heavy exchanged particles),
i.e. values in line with typical hadronic scales.

Parameters of the pole diagrams (bare masses
and coupling constants) are given in table \ref{pole_param}.
Note that the bare nucleon mass and bare $\pi N$ 
coupling constant $f^B_{\pi NN}$ are not free parameters,
because they are fixed by the physical values
of these quantities (cf. Ref. \cite{Schuetz94}). 
However, the cutoff at the 
$\pi NN$ vertex was allowed to vary, in order to 
fit the $P_{11}$ partial wave. The resulting parameters
for the nucleon pole are:
\be
M_0=1239 \  {\mathrm MeV},  \ \ \frac{(f^B_{NN\pi})^2}{4\pi}=0.0166,  \ \ \Lambda=1950 \ {\mathrm MeV}.
\ee
The cutoff masses for all other resonance diagrams were set to 2 GeV. 
Indeed the results depend only weakly on the
particular values of the cutoff masses, since their effects can be 
always compensated by a change in the corresponding coupling 
constants. The largeness of the cutoff masses in the resonance diagrams
is motivated by the specific analytical form of the employed resonance form 
factors, which fall off with momentum rather rapidly even for such a large 
cutoff mass \cite{Gasparyan02a}.

In general we adopt positive values for the sign of the bare coupling 
constants.  However, we use negative coupling constants if this leads
to a better agreement with the data. 
In the case of the $N^*_{P_{13}}(1720)\eta N$ vertex we changed
the sign of the coupling constant because that allow to
obtain a better description of the
$\pi N\to \eta N$ differential cross section
via an interference of the $P_{13}$ with other partial waves.
Finally, we would like to remark that 
among the three phase shift analyses
whose results are shown in figures we use the energy independent 
analysis from Ref.~\cite{SM95} as main guideline 
for the fitting procedure. 

\subsection{$\pi N$ elastic scattering}

We start the discussion of the elastic $\pi N$ data by first looking 
at the phase shifts as they
result from the original model of O. Krehl et al.~\cite{Krehl00}
(cf. the dashed curves in figs.~\ref{full1} and \ref{full3}).
In general, the quality of the description is
rather good, but there are some unsatisfactory 
features which we would like to point out here.

First, there are significant deviations of the
model results from the data in some partial waves,
specifically in the $P_{13}$, $S_{31}$, and $D_{33}$ waves.
Evidently, the discrepancies are primarily due to the
presence of resonances in these partial waves, which
are not yet included in the model. However, it is easy to
see, that the inclusion of the resonances in question alone 
will not help in the case of $P_{13}$ and $S_{31}$. This
is because such resonance contributions will  
vanish again above the position of the resonance 
within the energy range given roughly by the width 
and the phase shift will change by $180^\circ$ (if the
resonance contribution and the background have the same signs)
or turn back to the background (if they have opposite signs).
However, as one can see from figs.~\ref{full1} and \ref{full3},
in the $P_{13}$ partial wave the phase goes 
in opposite direction to the data, and in
$S_{31}$ partial wave the deviation from the
data at energies above the position of
the resonance is huge. 

The second problem of the $\pi N$ model of Krehl et al. 
is the presence of a long tail of the $S_{11}(1650)$ resonance.
This leads to the undesirable feature, that even at very low energies the $S_{11}$ 
phase shift is strongly influenced by this resonance -- in conflict with
chiral symmetry. As
was shown by Weinberg and Tomozawa~\cite{Weinberg66,Tomozawa66}
the isovector $s$-wave $\pi N$ scattering length
is fully determined to leading order by the pion mass $m_\pi$
and pion decay constant $F_\pi$. Therefore, the presence of contributions
related to the $N^*(1650)$ resonance at low energies is unnatural 
and physically hard to justify. 
Fig.~\ref{S11} shows the discussed effect. One can see that 
the low energy $S_{11}$ phase shift even changes
its sign when all couplings to the $S_{11}(1650)$ resonance
are switched off. 

A detailed inspection of this problem revealed that
the long tail of the $S_{11}(1650)$ resonance is predominantly 
due to the rather hard form factors used in the model of Krehl
et al. and, in particular, in those diagrams contributing to 
the $\pi N\to \rho N$ transition potential
(there is a direct coupling of the $\rho N$ channel to the $S_{11}(1650)$). 
As was mentioned before, in the new model we want to avoid the use
of extremely large cutoff masses anyway. A further reduction of 
the near threshold contribution from the $S_{11}(1650)$ resonance 
is achieved by choosing the derivative coupling for 
the $N^*_{S_{11}}N\pi$ vertex (see appendix A) in analogy with 
the $NN\pi$ coupling. In the new model the missing strength
at low energies is
provided by the correlated $\pi\pi$-exchange in the $\sigma$-channel. 
It can be generated by allowing the subtraction constant, which occurs in
the corresponding dispersion relations, cf. Eq. (31) in Ref. \cite{Schuetz94},
and which was set to zero by hand in the old model \cite{Krehl00},
to assume a finite but still small value. 

Now, let us consider the non-resonant part (or the background) of the new model.
First one should note that the main contribution 
to the background at low energies is, of course, provided by diagrams
that involve only the $\pi N$ channel. Therefore, we start by discussing 
the importance of various $\pi N$ graphs for the different partial
waves. There are five diagrams in the $\pi N\to\pi N$ potential, cf. 
Fig. \ref{pingraf}a-e: correlated $\pi\pi$ exchange in the 
$J=0$, $I=0$ ($\sigma$) and $J=1$, $I=1$ ($\rho$) channels, 
nucleon and $\Delta$ $u$-channel
exchanges, and the nucleon ($s$-channel) pole diagram. 
It turned out that the contribution from the
$\Delta$-exchange is very small in all partial waves. 
(As a consequence of that, we do not include 
$u$-channel graphs involving heavier resonances!) 
The $S$ waves are dominated by the $\rho$ and $\sigma$ exchanges. 
The nucleon exchange becomes important in higher partial waves.
Note also, that the $\rho$-exchange alone provides such a strong attraction
in the $P_{11}$ partial wave that it is almost sufficient for
the formation of a resonance. However, 
it is partly cancelled by the contribution 
from the nucleon pole. One should emphasize here
that, in contrast to the
old model, we do not have much freedom in varying the strength of 
the $\rho$ and $\sigma$ exchanges 
(except for the subtraction constant mentioned above) since their contributions
at low energies are basically fixed due to our
choice of the form factors (see section II).
Thus, the simultaneous description of the background 
in seven partial waves with a rather 
small number of parameters is to be considered as a 
success of our model (cf. dash-dotted lines in fig.~\ref{full1} and fig.~\ref{full3}).
We have not included the $P_{11}$ partial wave in these consideration, 
because there the coupling to the $\sigma N$ channel plays
a rather important role.

As a confirmation for the quality of the background contribution we also looked
at the phase shifts with $J = 5/2$, cf. Fig.~\ref{phase5}. 
Those partial waves were not include in the fitting procedure and, therefore, are
genuine predictions of our model. It is evident that our results are quite in
line with the general trend of the data (disregarding the resonance structures,
of course). 

The next step is the inclusion of the inelastic channels.
The most important ones are those that represent effectively 
the $\pi\pi N$ channel, namely $\rho N$, $\sigma N$ and $\pi \Delta$.
The $\sigma N$ channel couples dominantly to the $P_{11}$ $\pi N$ partial wave.
It is a consequence of the parity difference between $\pi$ and $\sigma$, which implies
that the $P_{11}$ $\pi N$ partial wave couples to an $S$ wave in the $\sigma N$ 
system.
In the course of adjusting the free parameters attraction is introduced 
into the $\sigma N$ channel and also a strong $\pi N\to\sigma N$
transition  potential results. This, in turn, provides additional attraction
in the $\pi N$ channel via coupled-channels effects and eventually leads
to a dynamical generation of the $N^*(1440)$ (Roper) resonance in the
$P_{11}$ partial wave.  
This mechanism and also its implications for the Roper resonance were 
discussed extensively in previous studies \cite{Schuetz98,Krehl00} by the J\"ulich
group and therefore we do not repeat the arguments here. However, it is
certainly reassuring that also within the new model the Roper resonance turns out
to be dynamically generated and no genuine $N^*(1440)$ (three quark) resonance is needed
in order to explain the $P_{11}$ partial wave. 
  
The channels $\rho N$ and $\pi \Delta$ are important 
for the inelasticities at high energies in all partial waves, 
but in particular in the $D_{13}$, $P_{31}$, and $P_{33}$. In the
$P_{33}$ partial wave there are no resonances in this energy region 
that couple strongly to the $\pi N$ system \cite{Hagiwara02}.
Thus, coupling to those channels via $t$- and $u$-channel exchange
diagrams is the only source of inelasticity in the $P_{33}$ $\pi N$ partial wave. 
The most important diagrams for the $P_{33}$
inelasticity are the $\rho$-exchange in the $\pi N\to \pi\Delta$ potential and, partly,
the nucleon exchange in the $\pi N\to \rho N$ potential. 
One should mention here also the $\pi$-exchange diagram in the
$\pi N\to \rho N$ transition. It turns out to be much too 
strong in the $P_{13}$ and $S_{11}$ partial waves, independently of the cutoff used.
Its contribution alone produces a very strong cusp in the region of
the $\rho  N$ threshold in the $S_{11}$ phase shift, and drastically modifies the behavior 
of the $P_{13}$ phase shift, bending it upwards. 
Luckily the $\pi$-exchange contribution is
cancelled to a large extent by the $\pi N\to\rho N$ contact term from the Wess-Zumino 
Lagrangian, and also by the $\omega$-exchange diagram.
Ultimately, on the whole the phase shifts are not too much affected 
by the inelastic channels.

The final step in the description of the 
elastic $\pi N$ data consists in adding the resonance terms.
We included resonances in all partial waves except for the 
$P_{11}$ where our model reproduces the phase shift and inelasticity, 
including the structure associated with the Roper resonance, 
dynamically via a strong coupling to the $\sigma N$ channel, as mentioned 
already above.  
In the $S_{11}$ partial wave there are two resonances, namely the
$N^*(1535)$ and the $N^*(1650)$. The former dominates
the near threshold $\pi N\to \eta N$ 
cross section. (The $\eta N$ channel will be discussed in detail in the next 
subsection.) As can be seen from the parameters given in
table \ref{pole_param}, among the effective $\pi\pi N$ channels, 
the $\pi \Delta$ channel is allowed to couple to most of the resonances.
This channel becomes relevant already at rather
low energies (in contrast to the $\rho N$ channel) and it can contribute to
both ($I=1/2$ and $I=3/2$) isospin states. Since
we cannot calculate $\pi N\to \pi\pi N$ observables directly at the moment
(due to technical difficulties that arise from 3-body singularities)
-- which would allow to further constrain the relative importance
of the different $\pi\pi N$ channels --
we choose this particular channel for describing the bulk of the 
$\pi\pi N$ part of the $\pi N$ inelasticity. However, in addition the 
$\rho N$ channel needs to be coupled to some resonances namely
to $D_{13}(1520)$, $S_{11}(1650)$, and $D_{33}(1700)$. In those
cases the different energy behavior resulting from the $\rho N$ channel
is required for a satisfactory description of the experimental phase shifts
as well as the inelasticities. 

The position of the $P_{31}(1910)$ resonance is located already 
above the energy region we are interested in
(which is from $\pi N$ threshold up to $\sim1.9$ GeV).
Nevertheless it was included because 
its tail still influences noticeably the energy region around 
1.8 $\sim$ 1.9 GeV.

Note that, among others, the inelasticity in the $P_{13}$
partial wave shows an incorrect trend at higher energies, and
the data are underestimated. A similar, but less pronounced 
deficiency can be found in the $D_{13}$ inelasticity. Some authors claim, 
that there is a sizable contribution from the $\omega N$ channel,
which opens at around $1.7$ GeV, to these
particular partial waves \cite{Penner02}.
Therefore, the inclusion of the $\omega N$ channel might improve the 
description of these data.

Finally, let us mention that also the low
energy parameters of $\pi N$ scattering are in reasonable
agreement with available data, as it should be,
since we fit our model to the phase shift analyses.
The $S$ and $P$-wave scattering lengths and
volumes are collected in table \ref{scatlength}.

\subsection{Description of the $\eta N$ channel}

The reaction $\pi N\to \eta N$ near the $\eta N$ threshold
is closely related to the properties of the $N^*(1535)$ resonance.
The total cross section of this reaction has a very pronounced
peak structure at the position of the resonance (cf. Fig.~\ref{xseta}).
In the previous version of the J\"ulich $\pi N$ model
the total $\pi^-p \to \eta n$ cross section was overestimated 
by about 20-30\% around the maximum.  
The reason for this deficiency
is that only the $\pi N$ and $\eta N$ channels were 
allowed to couple to the $N^*(1535)$ resonance. Therefore, in 
order to describe the $S_{11}$ $\pi N$ amplitude
one had to generate basically the whole inelasticity in this partial wave 
by the coupling to the $\eta N$ channel. Indeed the contribution of the
$S_{11}$ partial wave to the inelastic $\pi^-p$ cross 
section is given by 
\be
\sigma_{in}=\frac{2\pi}{3k_1^2}(1-\eta^2),
\ee
which amounts to $\sigma_{in}\sim3.5$ mb at the maximum using the 
inelasticity $\eta$ as given by the phase shift analysis. 
However, the experimental $\pi^-p \to \eta n$ cross 
section is always below 3 mb, cf. Fig.~\ref{xseta}.
Thus, it is clear that there must be contributions of other 
channels to the $S_{11}$ inelasticity. 
The only other channel which is open 
at energies around the $\eta$ threshold is the $\pi\pi N$
channel. Indeed the $\pi\pi N$ channel
was found to be important in an analysis of the $\pi N$ S-waves
within the chiral unitary approach of Inoue et al. \cite{Inoue02}. 
Accordingly, 
we introduce a coupling of the $\pi \Delta$ system 
-- which in our model is one of the effective channels that
represent the $\pi\pi N$ channel --
to the $N^*(1535)$ resonance. This enables us 
to describe simultaneously the total $\pi^-p \to \eta n$ cross section
and the inelasticity in the $S_{11}$ partial wave in the
resonance region, as can be seen in Figs.~\ref{xseta} and \ref{full1},
respectively. 

The inclusion of an $N^*\Delta\pi$ coupling improves also
the description of the $S_{11}$ inelasticity above
the position of the $N^*(1535)$ resonance.
In Fig.~\ref{full1} one can see that the old J\"ulich model produces a 
strong dip in the $S_{11}$ inelasticity, which then leads to 
a similar dip in the $S$-wave $\pi^-p \to \eta n$
cross section. We found that the origin of this behavior is essentially a 
unitarity constraint from the $\pi N$ channel.
It can be easily understood schematically, if we
assume a two-channel problem involving only the $\pi N$ and $\eta N$ 
systems. We also assume that, 
apart from the $N^*(1535)$ resonance (whose contribution
drops quickly when one moves away from its peak)
there is some background contribution to the $\pi N\to \eta N$
transition potential and that at the same time 
(which is the crucial point) 
the direct $\eta N\to \eta N$ potential is negligibly small.
(These conditions are satisfied in the old J\"ulich 
model.) Then the $\pi N\to \eta N$ $T$-matrix
(we consider only the $S_{11}$ partial wave) is given by
\be
T_{\pi N\to\eta N}=V_{\pi N\to\eta N}(1 + G_0T_{\pi N\to\pi  N}),
\ee
which can be re-expressed in the form (see, e.g., Ref.~\cite{Hanhart99})
\be
T_{\pi N\to\eta N}=(1 + (\beta +ik_{\pi N})\frac{\eta e^{2i\delta}-1}{2ik_{\pi N}})
V_{\pi N\to\eta N}\, .
\label{beta}
\ee
Here 
$\beta$ is the inverse of the characteristic range of interaction, which
is determined by the principal value integral, $\delta$ and $\eta$
are the $S_{11}$ phase shift and inelasticity parameter and $k_{\pi N}$ is
the on-shell momentum in the $\pi N$ channel .
Let us now examine under what circumstances we can have
$T_{\pi N\to\eta N}=0$. Given our simplifying model assumption,
the condition $T_{\pi N\to\eta N}=0$ implies that 
$\eta=1$, and consequently $\beta$ is purely real. Then, it is convenient to rewrite 
Eq.~(\ref{beta}) as
\be
T_{\pi N\to\eta N}=e^{i\delta}\sqrt{1+\beta^2/k_{\pi N}^2}\sin{(\gamma-\alpha)}
V_{\pi N\to\eta N},
\label{zero}
\ee
where $\gamma=\arctan{(\beta/k_{\pi N})}$ and $\alpha=\delta-\pi/2$.
Note that in the specific situation we discuss the phase $\delta$ crosses $\pi/2$ ($\alpha=0$),
due to the presence of the $N^*(1650)$ resonance in the $\pi N\to \pi N$
interaction, and then 
continues to rise rapidly, whereas $\beta$ is a smooth
function of $k_{\pi N}$ and has a typical value in the order of 
several hundreds MeV (the exact value is, of course, model dependent),
so that in the region of interest we have $\gamma\lesssim 1$.
It is thus easy to convince oneself that the expression in Eq.~(\ref{zero})
equals zero at some energy above (but not far from) the position 
of the $N^*(1650)$ resonance. Expanding $T_{\pi N\to\eta N}$ in 
powers of $Z-Z_0$, where $Z_0$ is the position of the ``zero'',
one  can see that the $\pi^-p \to \eta n$ cross section 
is proportional to $(Z-Z_0)^2$ - which explains the structure
of the dip in the cross section exhibited by the old J\"ulich model (dotted line  
in Fig.~\ref{xseta}). It is interesting to note that the same effect
can be found in other model analyses , e.g. in the ones by 
Gridnev and Kozlenko~\cite{Gridnev99} and by the Giessen 
group~\cite{Penner02a,Shklyar03} 
In general, when there are more than two channels, 
$\beta$ becomes complex and the cross section at the dip will be finite -- 
but it will be still small
(provided that the inelasticity is not too large).

In our model the $S_{11}$ inelasticity in the
energy region around the $\eta$ threshold is partly determined by the
$N^*(1535)\Delta\pi$ coupling. In the $\Delta\pi$ system this resonance
couples to a pure $D$-wave. Because of that the 
maximum of the $\pi N\to \pi \Delta $ transition cross section is shifted 
to somewhat higher energies as compared to the resonance energy. Its
contribution to the inelasticity is likewise shifted to somewhat higher energies
and fills up the dip that can be seen in the $S_{11}$ inelasticity predicted
by the old model, cf. Fig.~\ref{full1}. It also smoothens out the effect
discussed above and, therefore, we can achieve a fairly realistic description
of the energy dependence of $\sigma_{\pi^-p \to \eta n}$ over the region of the
$N^*(1650)$ resonance. Specifically, we don't get this strong double hump
structure prominently visible in the model analysis of Ref. \cite{Penner02a},
cf. their Fig. 7.

Let us now look at the energy dependence of the total cross
section over a wider energy range and also at the 
$\pi^- p\to \eta n$ differential cross section 
in order to examine the importance of higher partial waves.
To include the effect of higher partial 
waves we introduced a coupling of the $\eta N$ system 
to the $P_{13}(1720)$ and $D_{13}(1520)$ resonances.
Those are the most pronounced resonances in
the energy region below $1.9$ GeV
that couple strongly to the $\pi N$ system.
Note that there are other 3-star $N^*$ resonances \cite{Hagiwara02}
in this region. However, we do not include those because their
coupling to the $\pi N$ channel is very weak and therefore their
parameters cannot be sufficiently constrained from the $\pi N$ data.

At energies below $1.6$ GeV the slight deviation
of the differential cross section from 
the isotropic distribution can be easily described
by the interference of the $D_{13}$ resonance with the 
$S$-wave amplitude \cite{Krehl98a}, cf. fig.~\ref{dwq1}. For the total 
cross section the $D_{13}$ contribution is of minor importance. 
Above $1.6$ GeV the total cross section
can be described by introducing a coupling of the $\eta N$ system
to the $P_{13}(1720)$ resonance as is evidenced by the results shown
in fig.~\ref{xseta}. However, as is obvious 
from fig.~\ref{dwq2}, this coupling alone is not sufficient
to achieve also good agreement with the data for the 
differential cross section in this energy region.
Most likely that points to missing contributions from higher
partial waves, and specifically from $J=5/2$ resonances. 
At present we do not aim to include those. We 
would like to remark also that the existing data do not allow one 
to discriminate between different partial-wave
contributions -- one would need to know polarization observables
for this purpose.

Finally, we want to draw attention to the fact 
that in our model there is also a background contribution to the 
$\pi N \to \eta N$ transition interaction 
which is provided by $t$-channel exchange of the $a_0$(980) meson, 
cf. Fig.~\ref{etangraf}b. However, the role of $a_0$ exchange is now strongly reduced 
as compared to the old J\"ulich model, mainly because in the present model we avoid 
large values of the cutoff mass. 
In any case, the influence of the $a_0$(980) meson is
suppressed at energies above the $N^*(1535)$
resonance due to the mechanism discussed above.

We also want to present the $\eta N$ effective range parameters
predicted by our model. They are 
\be
\nonumber
a_{\eta N}&=&(0.41+i0.26) fm, \\ 
r_{\eta N}&=&(-3.4+i0.4) fm.
\ee
Obviously, our result for Re($a_{\eta N}$) is at the lower end
of the spectrum of values that one can find in the literature,
cf., e.g., the compilation given in Table 1 of Ref. \cite{Sibirtsev02}.
In fact, it is even slightly lower than the one of the old J\"ulich
model, which yields $a_{\eta N}$ = (0.42+$i$0.34) fm. However, we want to emphasize 
that such a value is pretty much in line with conclusions drawn from recent analyses 
of the $\eta N$ final state interaction in the 
reactions $\gamma d\to np\eta$~\cite{Sibirtsev02} 
and $pn\to d\eta$~\cite{Grishina99,Garcilazo02}.
 
\section{Summary}\label{summary}

We have presented results of 
an extended and improved version of the J\"ulich $\pi N$ model.
The model is based on the meson exchange picture and it is derived
in its main part from the phenomenological Wess-Zumino Lagrangian,
consistent with chiral symmetry.
The $\pi N$ interaction in the scalar-isoscalar 
and vector-isovector channels is calculated
by means of dispersion relations from correlated $\pi\pi$ exchange
in order to constrain the contributions of the corresponding
$\sigma$ and $\rho$ exchanges. In the present work ambiguities
in the treatment of dispersion relations (cf. sect. IIIB of Ref.
\cite{Schuetz94}) are even further reduced  
by a choice of the form factors that does not 
modify the strength and $t$ dependence of the interaction at low energies.
In addition some more improvements have been implemented. 
In particular, we now use derivative coupling for the $S_{11}$ $N^*$ resonances 
at the $\pi N$ and $\eta N$ vertices, as is demanded by chiral symmetry anyway. 
We also include some more resonance diagrams, specifically for the 
$S_{13}$(1620), $P_{13}$(1720), $P_{13}$(1910), $D_{13}$(1520), and
$D_{33}$(1700) resonances. 

The potential constructed in this way was unitarized
in a coupled channels Lippmann-Schwinger equation to obtain the 
reaction amplitudes for various processes.
The reaction channels included in the present investigation are
$\pi N$, $\eta N$, $\sigma N$, $\rho N$, and $\pi\Delta$, 
where the latter three channels are understood 
as an effective description of the physical $\pi \pi N$ state.

With the new model an excellent quantitative reproduction 
of the $\pi N$ phase shifts and inelasticity parameters 
in the energy region up to 1.9 GeV and for total angular momenta 
$J\leq3/2$ was achieved. 
In addition a good description of the background in the $J=5/2$ 
partial waves was obtained automatically.
As far as the $P_{11}$ partial wave is concerned we confirm the
results of our earlier investigations that the structure associated with 
the Roper resonance is generated dynamically by the model so that no
genuine $N^*$(1440) (three quark) resonance diagram needs to be included. 

As the main new aspect we studied in detail the coupling of the $\pi N$ system 
to the $\eta N$ channel.  First of all we showed that the overestimation
of the $\pi^- p\to \eta n$ cross section 
in the region of the $N^*(1535)$ resonance  
by the old J\"ulich model can be removed by introducing 
additional flux from the $\pi N$ to the $\pi \Delta$ channel.
Furthermore, the inclusion of the 
$\eta N$ coupling to the $P_{13}(1720)$ resonance
turned out to lead to a significant improvement
of the total cross section at higher energies.
At the same time the puzzle of the dip
in the $S_{11}$ inelasticity, present in the old J\"ulich model but 
also in other models in the literature \cite{Gridnev99,Penner02a}, 
could be explained. The origin 
of this deficiency turned out to be an almost model independent
effect of coupled-channels unitarity constraints.
We also improved the description of the
$\pi^- p\to \eta n$ differential cross section.
A remaining discrepancy with the data at higher energies 
is most likely caused by contributions from partial waves with $J > 3/2$ 
which are not included in our model calculation.
Note, that a detailed partial wave analysis of this reaction 
at such energies is presently impossible because of the lack of 
data on polarization observables.

The model in its present form enables a straightforward inclusion of
further reaction channels, and specifically those
nearest in energy namely $K \Lambda $, $K\Sigma$, and $\omega N$. 
Such an extension of the present model in this direction is planned
for the future.

\section*{Acknowledgments}
We thank S. Krewald and I.R. Afnan for stimulating discussions. 

\appendix 

\section{The pseudopotential}\label{ana}
%%%%%%%%%%%%%%%%%%%%%%%%%%%%%%%%%%%%%%%%%%%%%%%%%%%%%%%%%%%%%%%%%%
In this appendix we give the expressions for those contributions
to the interaction potentials that differ from our earlier 
work \cite{Krehl00}. For convenience we also summarize here all pole 
diagrams since most of them were not included in the old model. 
All other expressions for the pseudopotential
can be found in Appendix A of Ref. \cite{Krehl00}. 
The notation for the different particles
and their momenta is given in Fig. \ref{pingraf}.  
$E_1$, $E_3$, $\omega_2$ and $\omega_4$
indicate on--mass--shell--energies of baryons $1$ and $3$ and,
respectively those of
mesons $2$ und $4$:
\begin{equation}
E_i = \sqrt {m_i^2 + \vec p_i^{\, 2}} \;\;\; ; \;\;\;
\omega_i =  \sqrt {m_i^2 + \vec p_i^{\, 2}} \;\; .
\end{equation}
$q$ is the four-momentum of the intermediate particle. The tensor
operator $P^{\mu\nu}$ is given in Eq. (A12) of Ref. \cite{Krehl00}.

Since we work in time-ordered perturbation theory, all the potentials
contain the normalization factor \begin{equation}
\kappa = {1\over (2\pi)^3}\sqrt{{m_1\over E_1}{m_3\over E_3}}
\sqrt{1\over 2\omega_2 2\omega_4} \;\; .
\end{equation}

\subsection{$\npi \to \npi$}
\begin{figure}
\bc
\includegraphics[width=7cm]{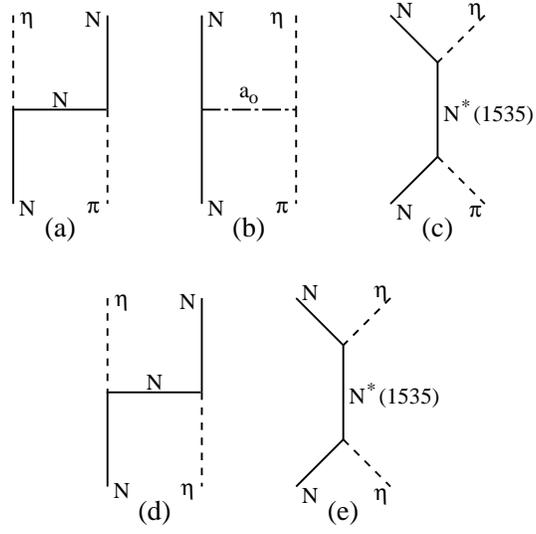}
\ec
\caption{Contribution to the elastic $\pi N$ channel.}
\label{pingraf}
\end{figure}

\begin{figure}
\bc
\includegraphics[width=7cm]{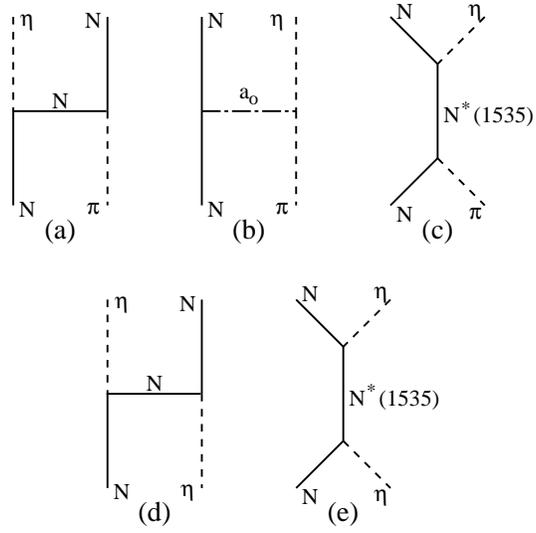}
\ec
\caption{Contribution to the $\pi N \to \eta N$ transition and to the 
$\eta N$ channel.}
\label{etangraf}
\end{figure}

\begin{figure}
\bc
\includegraphics[width=14cm]{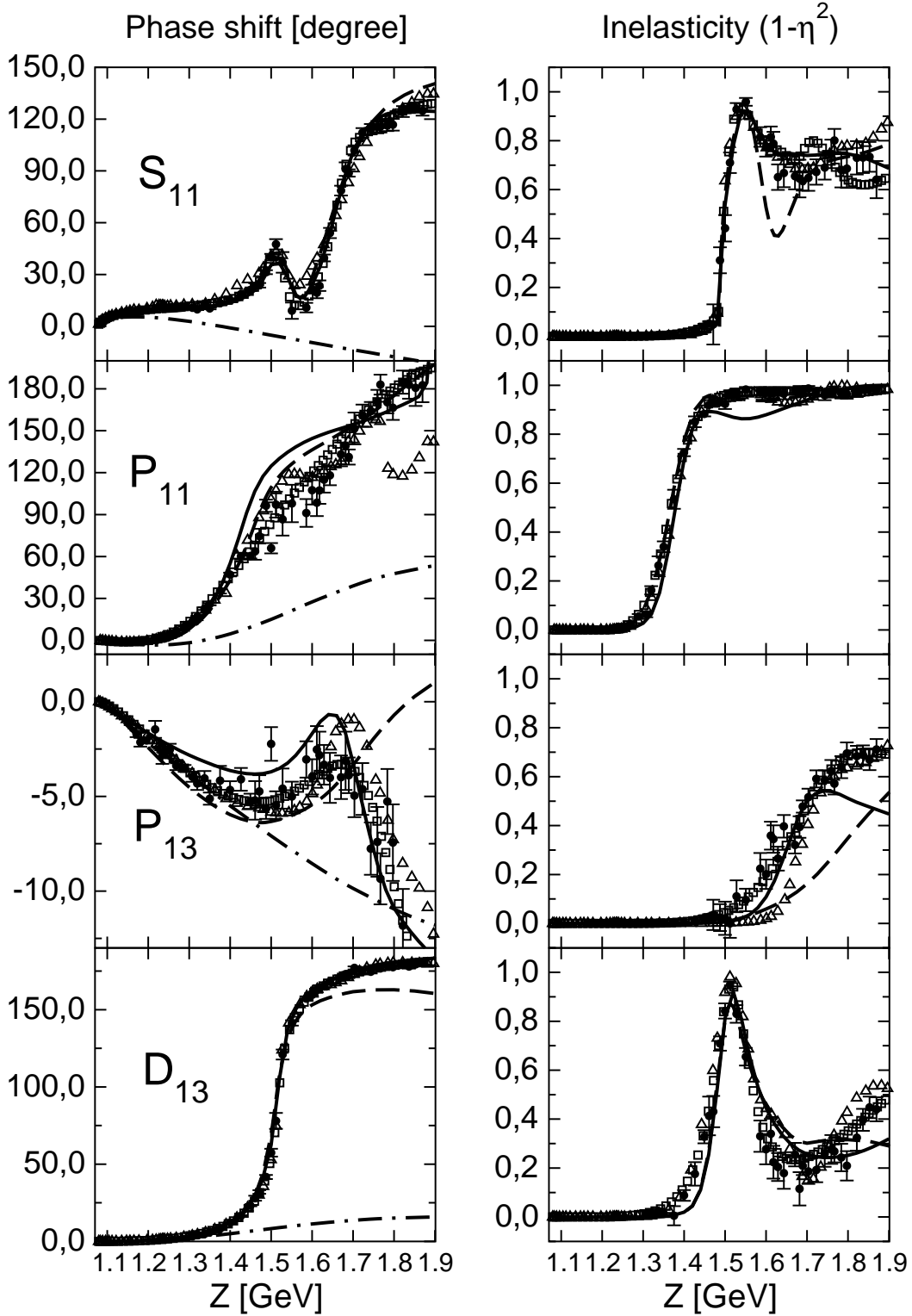}
\ec
\caption{The $\pi N$ phase shifts and
inelasticities for the isospin $I=1/2$ partial waves. 
The dashed curves show the results 
of the $\pi N$ model of O. Krehl et al. \protect\cite{Krehl00}. 
The dash-dotted curves represent the results based on
the background contributions of our new model, as discussed in 
the text. The results of the full model are given by the 
solid lines. The data are
from the phase shift analyses KA84\protect\cite{Koch86},
SM95\protect\cite{SM95}, and SE-SM95\protect\cite{SM95}.
}
\label{full1}
\end{figure}

\begin{figure}
\bc
\includegraphics[width=14cm]{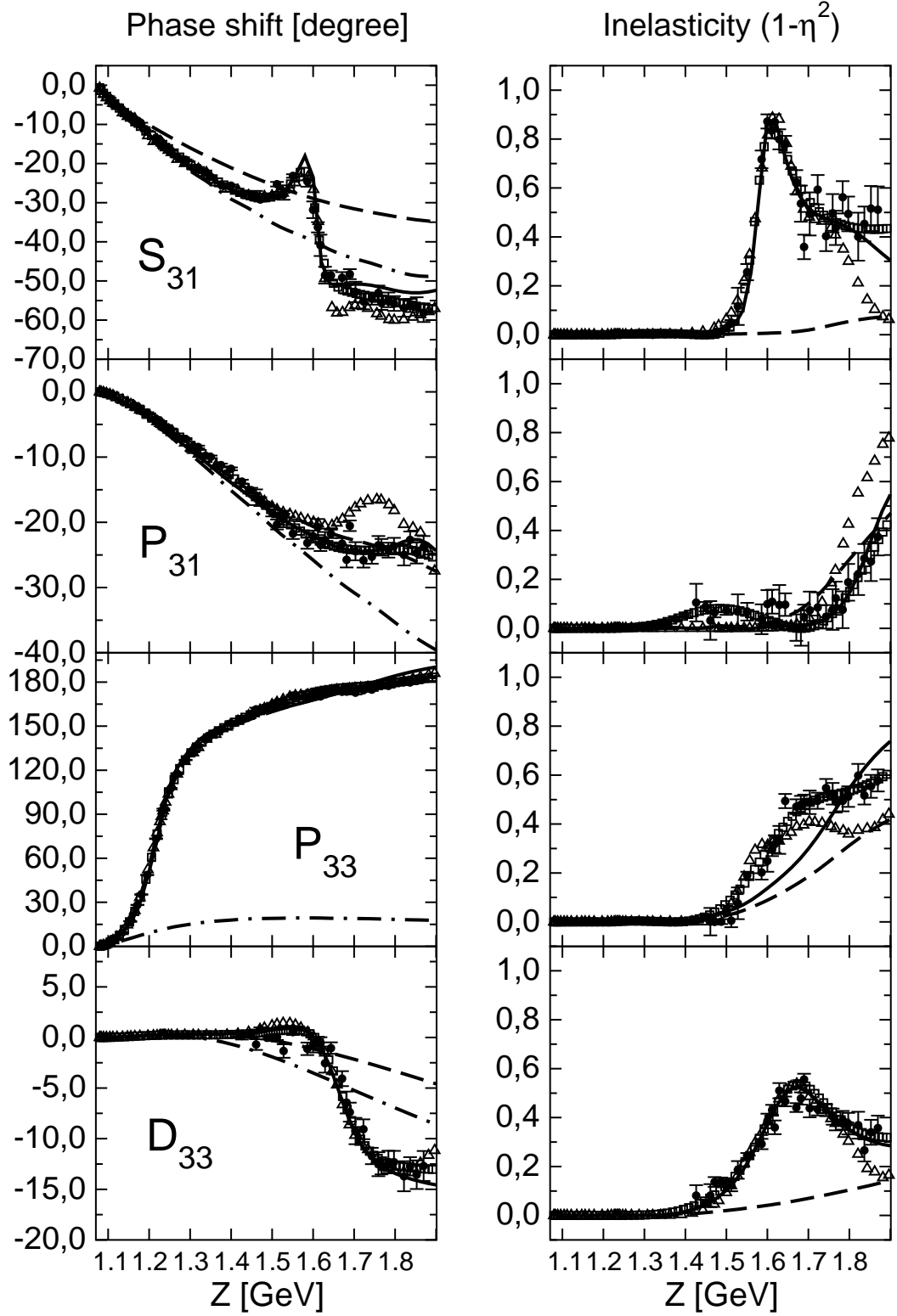}
\ec
\caption{The $\pi N$ phase shifts and
inelasticities for the isospin $I=3/2$ partial waves.
Same description of curves and experiments 
as in fig.~\ref{full1}.}
\label{full3}
\end{figure}

\begin{figure}
\bc
\vspace{0.5cm}
\includegraphics[width=14cm,height=20cm]{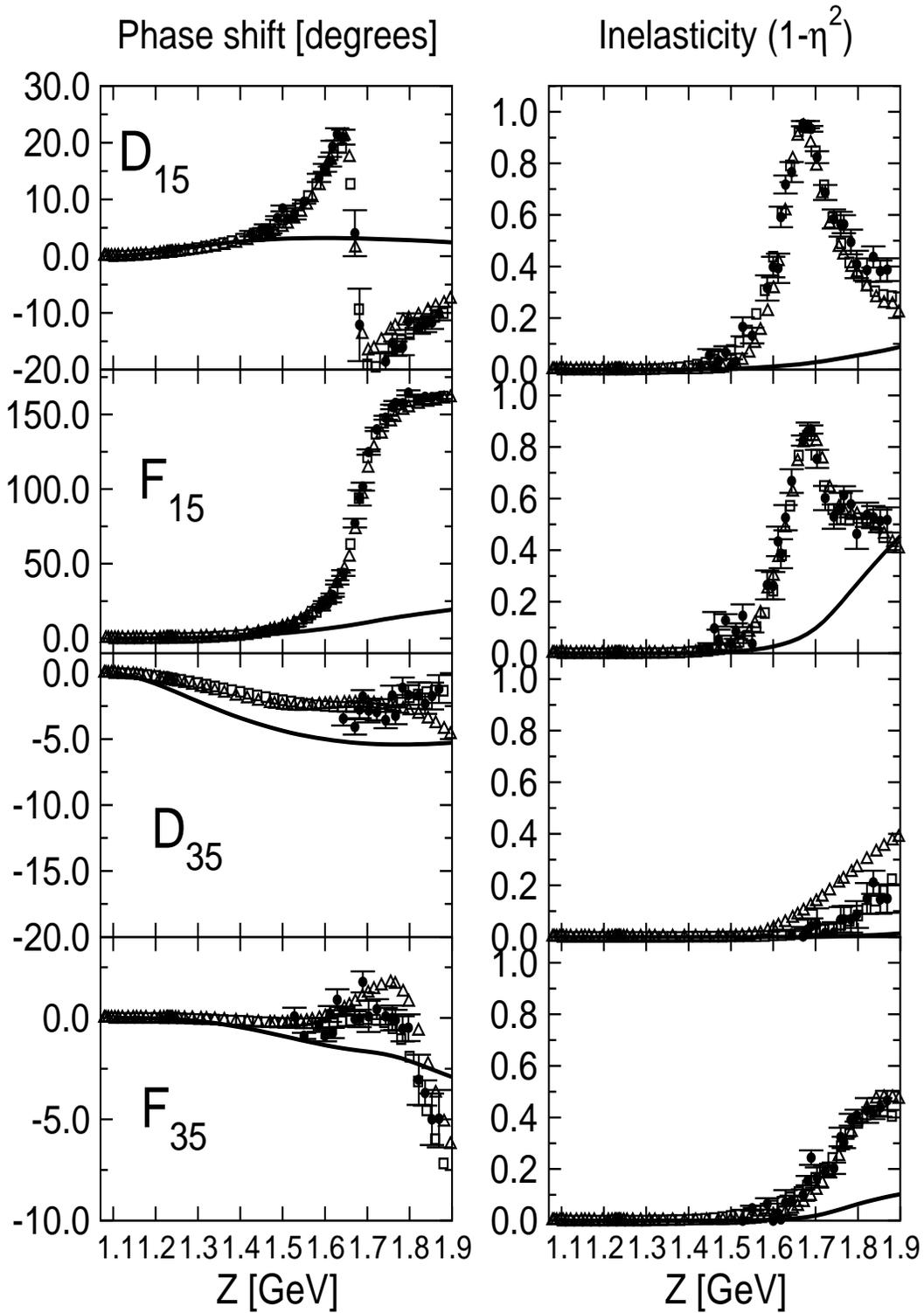}
\ec
\caption{Phase shifts and inelasticity for $\pi N$ partial waves with $J=5/2$. 
Same description of curves and experiments as in fig.~\ref{full1}.}
\label{phase5}
\end{figure}

\begin{figure}
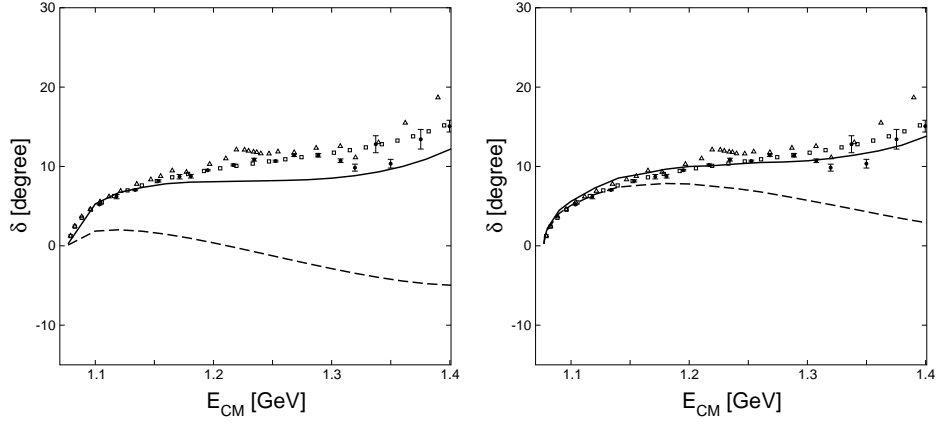

\bc
\includegraphics[width=6.0cm]{S11_1.4.eps} \hskip 0.2cm 
\includegraphics[width=6.0cm]{S11_new_1.4.eps}
\ec
\caption{$\pi N$ phase shift in the $S_{11}$ partial wave.
The results of the model of Krehl et al. \protect\cite{Krehl00} are
shown on the left side, those of the new model on the right side. 
The curves correspond to the full model (solid line) and to 
the full model with the contribution of the $S_{11}(1650)$ 
resonance switched off (dashed line).
The data are from the phase shift analyses 
KA84 \protect\cite{Koch86},
SM95 \protect\cite{SM95}, and SE-SM95 \protect\cite{SM95}.}
\label{S11}
\end{figure}

\begin{figure}
\bc
\vspace{0.5cm}
\includegraphics[width=14cm]{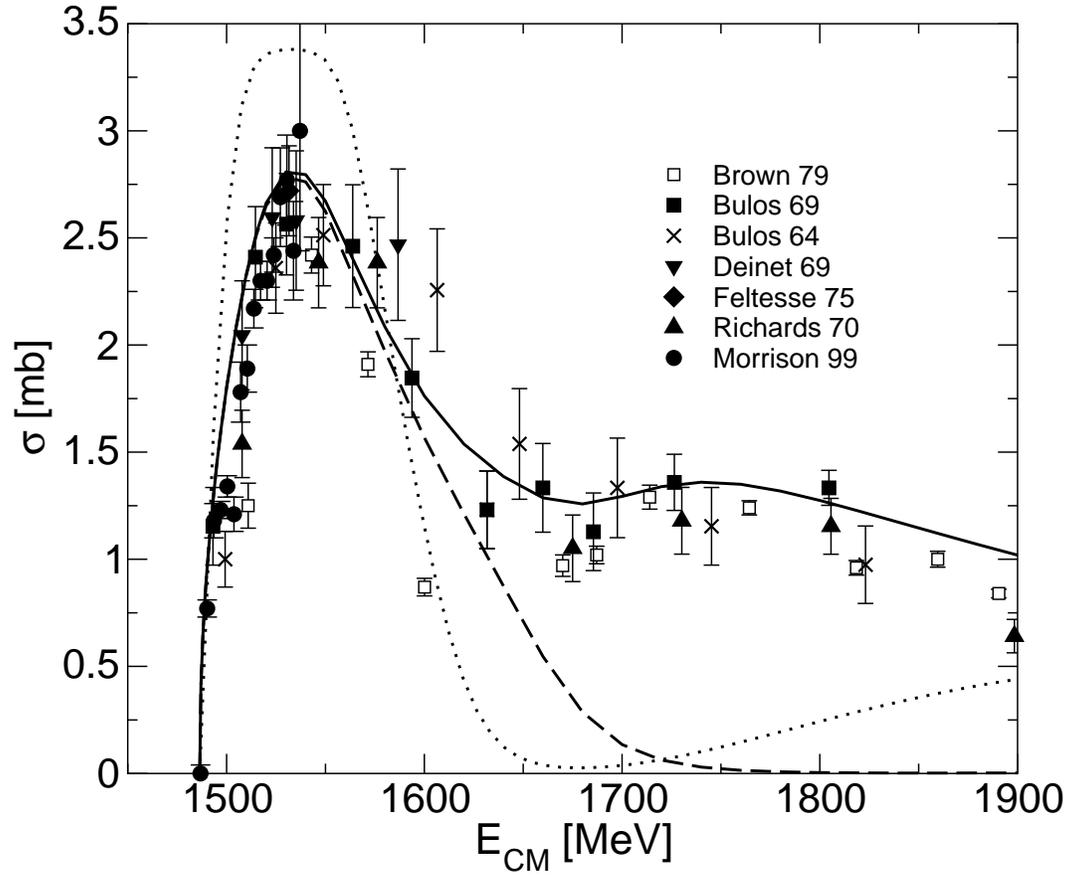}
\ec
\caption{$\pi^- p\to \eta n$ total cross section. 
The solid line corresponds to the full calculation.
The dashed line indicates the pure $s$-wave contribution.
The results of the old model are shown as 
a dotted line (only $s$-wave). 
The data are from Refs. \protect\cite{Brown79,Bulos64,Bulos69,Deinet69,
Feltesse75,Richards70,Morrison99}.}
\label{xseta}

\end{figure}

\begin{figure}
\bc
\includegraphics[width=14cm]{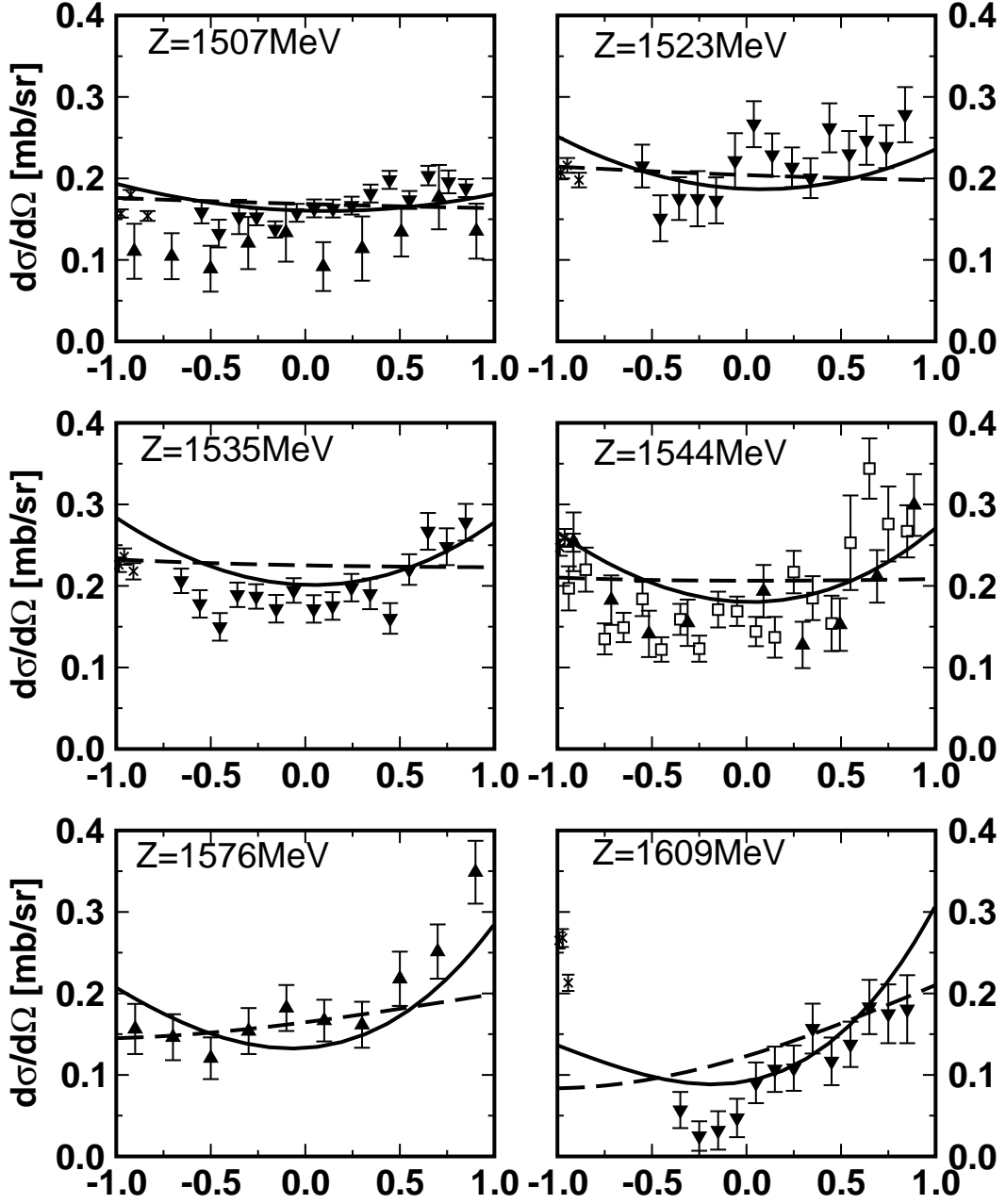}
\ec
\caption{$\pi^- N\to \eta n$ differential cross section at
energies close to the $\eta N$ threshold.
The solid curve corresponds to the 
full result. The dashed line corresponds to
the case when the $D_{13}$ partial wave
is switched off.
The data are from \cite{Brown79}($\square$),
\cite{Deinet69}($\blacktriangledown$),\cite{Richards70}($\blacktriangle$),
\cite{Debenham75}($\times$).}
\label{dwq1}
\end{figure}

\begin{figure}
\bc
\includegraphics[width=14cm]{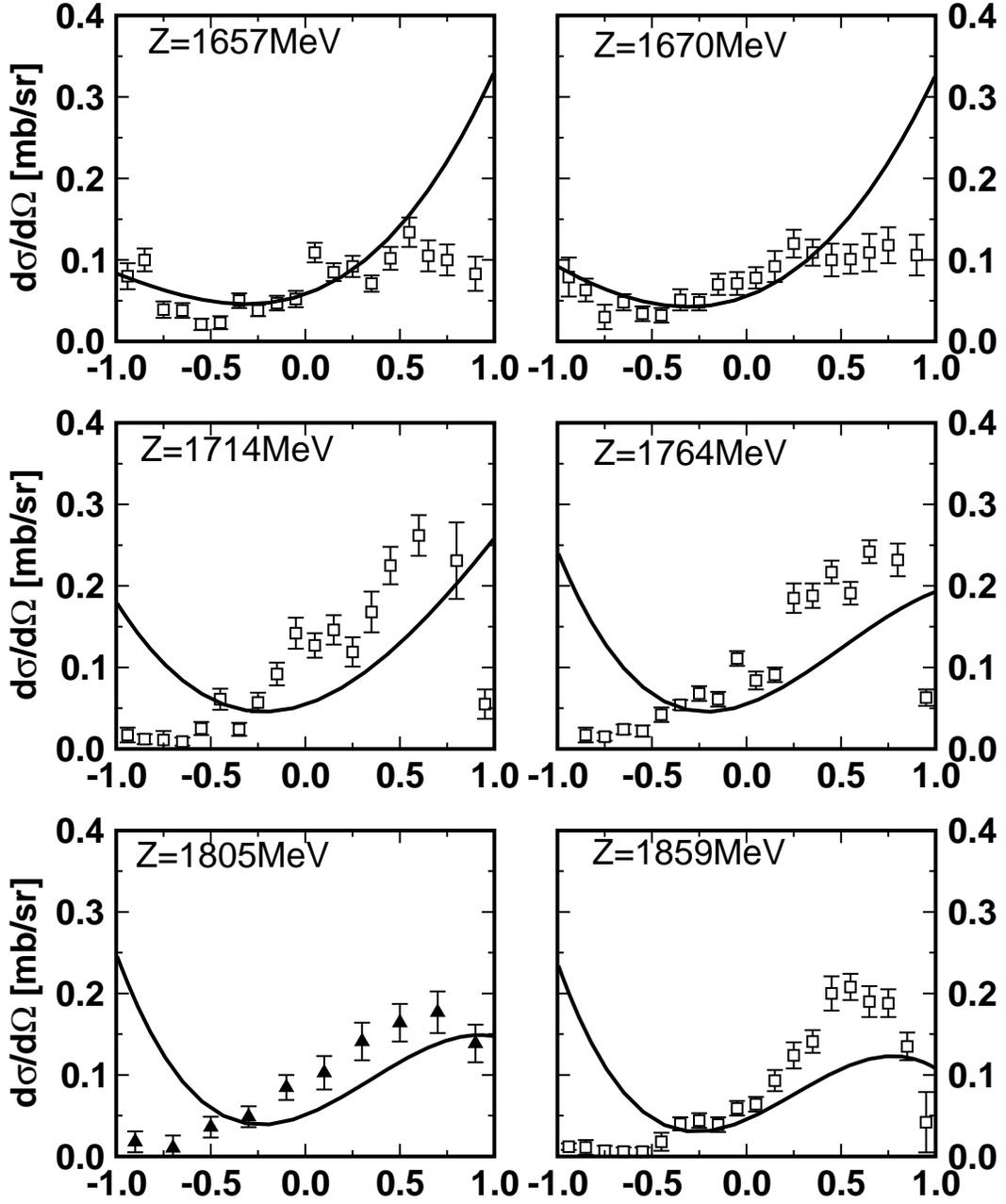}
\ec
\caption{$\pi^- N\to \eta n$ differential cross section at
higher energies.
The data are from \cite{Brown79}($\square$),
\cite{Deinet69}($\blacktriangledown$),\cite{Richards70}($\blacktriangle$),
\cite{Debenham75}($\times$).}
\label{dwq2}

\end{figure}

\begin{itemize}
\item correlated $\pi\pi$ exchange in the {$\sigma$} channel (Fig.
\ref{pingraf}c)\\
\begin{equation}
-\kappa  \bar u (\vec{p}_3,\lambda_3)
u(\vec{p}_1,\lambda_1)\left[ A_0 + 16(-2p_{2\mu}p_4^{\mu}) \int dt'
\frac{Im(f^0_+(t'))}{(t'-2m_{\pi}^2)(t'-4m_N^2)}
P(t')\right ] IF_{\sigma t}(I),
\end{equation}
where
$P(t')=\frac{1}{2\omega_{t'}}\left(\frac{1}{E-\omega_2-E_3-\omega_{t'}}+\frac{1}{E-\omega_4-E_1-\omega_{t'}}\right)$,
$\omega_{t'}=\sqrt{q^2+t'}$ and $f$ is a Frazer-Fulco amplitude
\cite{Frazer60,Krehl99}. The isospin  coefficients are equal to 
$IF_{\sigma t}(1/2)=1,IF_{\sigma t}(3/2)=1.$

\item $N^*(S_{11},S_{31})$ pole diagrams (Fig. \ref{pingraf}g)
\begin{equation}
\kappa \frac{f^2_{N^*N\pi}}{m_\pi^2}
\bar u (\vec{p}_3,\lambda_3)p\slas_4
\frac{1}{2m^0_{N^*}}
\frac{q\sla+m^0_{N^*}}{E-m^0_{N^*}}
p\slas_2 u(\vec{p}_1,\lambda_1) IF_{N^*s}(I).
\end{equation}

\item Nucleon, $N^*(P_{31})$ pole diagrams
\begin{equation}
\kappa \frac{f^2_{N^*N\pi}}{m_\pi^2}
\bar u (\vec{p}_3,\lambda_3) \gamma^5 p\slas_4
\frac{1}{2m^0_{N^*}}
\frac{q\sla+m^0_{N^*}}{E-m^0_{N^*}}
\gamma^5 p\slas_2 u(\vec{p}_1,\lambda_1) IF_{N^*s}(I).
\end{equation}

\item $N^*(P_{13},P_{33})$ pole diagrams
\begin{equation}
\kappa \frac{f^2_{N^*N\pi}}{m_\pi^2}
\bar u (\vec{p}_3,\lambda_3)p_{4\mu}
\frac{1}{2m^0_{N^*}}
\frac{P^{\mu\nu}(q)}{E-m^0_{N^*}}
p_{2\nu} u(\vec{p}_1,\lambda_1) IF_{N^*s}(I).
\end{equation}

\item $N^*(D_{13},D_{33})$ pole diagrams
\begin{equation}
\kappa \frac{f^2_{N^*N\pi}}{m_\pi^4}
\bar u (\vec{p}_3,\lambda_3) \gamma^5 p\slas_4 p_{4\mu}
\frac{1}{2m^0_{N^*}}
\frac{P^{\mu\nu}(q)}{E-m^0_{N^*}}
\gamma^5 p\slas_2 p_{2\nu} u(\vec{p}_1,\lambda_1) IF_{N^*s}(I).
\end{equation}
$IF_{N^*s}(1/2)=3,IF_{N^*s}(3/2)=1.$

\subsection{$\npi \to \neta$}
\item $N^*(S_{11})$ pole diagram
\begin{equation}
\kappa \frac{f_{N^*N\pi} f_{N^*N\eta}}{m_\pi^2}
\bar u (\vec{p}_3,\lambda_3)p\slas_4
\frac{1}{2m^0_{N^*}}
\frac{q\sla+m^0_{N^*}}{E-m^0_{N^*}}
p\slas_2 u(\vec{p}_1,\lambda_1) IF_{N^*s}(I).
\end{equation}

\item $N^*(P_{13})$ pole diagram
\begin{equation}
\kappa \frac{f_{N^*N\pi}f_{N^*N\eta}}{m_\pi^2}
\bar u (\vec{p}_3,\lambda_3)p_{4\mu}
\frac{1}{2m^0_{N^*}}
\frac{P^{\mu\nu}(q)}{E-m^0_{N^*}}
p_{2\nu} u(\vec{p}_1,\lambda_1) IF_{N^*s}(I).
\end{equation}

\item $N^*(D_{13})$ pole diagram
\begin{equation}
\kappa \frac{f_{N^*N\pi} f_{N^*N\eta}}{m_\pi^4}
\bar u (\vec{p}_3,\lambda_3) \gamma^5 p\slas_4 p_{4\mu}
\frac{1}{2m^0_{N^*}}
\frac{P^{\mu\nu}(q)}{E-m^0_{N^*}}
\gamma^5 p\slas_2 p_{2\nu} u(\vec{p}_1,\lambda_1) IF_{N^*s}(I).
\end{equation}
$IF_{N^*s}(1/2)=\sqrt{3}.$

\subsection{$\npi \to \nrho$}
\item $N^*(S_{11})$ pole diagram
\begin{equation}
-i \kappa \frac{f_{N^*N\pi} g_{N^*N\rho}}{m_\pi}
\bar u (\vec{p}_3,\lambda_3)\gamma^5 \epsilon\sla^*(\vec{p}_4,\lambda_4)
\frac{1}{2m^0_{N^*}}
\frac{q\sla+m^0_{N^*}}{E-m^0_{N^*}}
p\slas_2 u(\vec{p}_1,\lambda_1) IF_{N^*s}(I).
\end{equation}

\item $N^*(D_{13},D_{33})$ pole diagrams
\be
\nonumber i\kappa\frac{f_{N^*N\pi}f_{N^*N\rho}}{m^2_{\pi}m_{\rho}}
\bar u (\vec{p}_3,\lambda_3) (p\slas_{4}\epsilon^*_\mu(\vec{p}_4,\lambda_4)-p_{4\mu}\epsilon\sla^*(\vec{p}_4,\lambda_4)) \\
\times \frac{P^{\mu\nu}(q)}{2m_{N^*}(E-m^0_{N^*})}
p_{2_\nu} \gamma^5 p\slas_2
u(\vec{p}_1,\lambda_1) IF_{N^*s}(I)
\ee
$IF_{N^*s}(1/2)=3,IF_{N^*s}(3/2)=1.$
\subsection{$\npi \to \dpi$}
\item $N^*(S_{11},S_{31})$ pole diagrams
\begin{equation}
\kappa\frac{f_{N^*N\pi}f_{N^*\Delta\pi}}{m^2_{\pi}}
\bar{u}_{\mu} (\vec{p}_3,\lambda_3) p_4^\mu \gamma^5
\frac{1}{2m^0_{N^*}}
\frac{q\sla+m^0_{N^*}}{E-m^0_{N^*}}
p\slas_2 u(\vec{p}_1,\lambda_1) IF_{N^*s}(I).
\end{equation}

\item $N^*(P_{31})$ pole diagram
\begin{equation}
\kappa\frac{f_{N^*N\pi}f_{N^*\Delta\pi}}{m^2_{\pi}}
\bar{u}_{\mu} (\vec{p}_3,\lambda_3) p_4^\mu
\frac{1}{2m^0_{N^*}}
\frac{q\sla+m^0_{N^*}}{E-m^0_{N^*}}
\gamma^5 p\slas_2 u(\vec{p}_1,\lambda_1) IF_{N^*s}(I).
\end{equation}

\item $N^*(P_{13})$ pole diagram
\begin{equation}
\kappa \frac{f_{N^*N\pi}f_{N^*\Delta\pi}}{m_\pi^2}
\bar{u}_{\mu} (\vec{p}_3,\lambda_3)\gamma^5 p\slas_4
\frac{1}{2m^0_{N^*}}
\frac{P^{\mu\nu}(q)}{E-m^0_{N^*}}
p_{2\nu} u(\vec{p}_1,\lambda_1) IF_{N^*s}(I).
\end{equation}

\item $N^*(D_{13},D_{33})$ pole diagrams
\begin{equation}
-\kappa \frac{f_{N^*N\pi} f_{N^*\Delta\pi}}{m_\pi^3}
\bar{u}_{\mu} (\vec{p}_3,\lambda_3) p\slas_4
\frac{1}{2m^0_{N^*}}
\frac{P^{\mu\nu}(q)}{E-m^0_{N^*}}
\gamma^5 p\slas_2 p_{2\nu} u(\vec{p}_1,\lambda_1) IF_{N^*s}(I).
\end{equation}
$IF_{N^*s}(1/2)=-\sqrt{6},IF_{N^*s}(3/2)=\sqrt{\frac53}.$

\subsection{$\neta \to \neta$}
\item $N^*(S_{11})$ pole diagram
\begin{equation}
\kappa \frac{f_{N^*N\eta}^2}{m_\pi^2}
\bar u (\vec{p}_3,\lambda_3)p\slas_4
\frac{1}{2m^0_{N^*}}
\frac{q\sla+m^0_{N^*}}{E-m^0_{N^*}}
p\slas_2 u(\vec{p}_1,\lambda_1) IF_{N^*s}(I).
\end{equation}

\item $N^*(P_{13})$ pole diagram
\begin{equation}
\kappa \frac{f_{N^*N\eta}^2}{m_\pi^2}
\bar u (\vec{p}_3,\lambda_3)p_{4\mu}
\frac{1}{2m^0_{N^*}}
\frac{P^{\mu\nu}(q)}{E-m^0_{N^*}}
p_{2\nu} u(\vec{p}_1,\lambda_1) IF_{N^*s}(I).
\end{equation}

\item $N^*(D_{13})$ pole diagram
\begin{equation}
\kappa \frac{f_{N^*N\eta}^2}{m_\pi^4}
\bar u (\vec{p}_3,\lambda_3) \gamma^5 p\slas_4 p_{4\mu}
\frac{1}{2m^0_{N^*}}
\frac{P^{\mu\nu}(q)}{E-m^0_{N^*}}
\gamma^5 p\slas_2 p_{2\nu} u(\vec{p}_1,\lambda_1) IF_{N^*s}(I).
\end{equation}
$IF_{N^*s}(1/2)=1.$

\subsection{$\neta \to \nrho$}
\item $N^*(S_{11})$ pole diagram
\begin{equation}
-i \kappa \frac{f_{N^*N\eta} g_{N^*N\rho}}{m_\pi}
\bar u (\vec{p}_3,\lambda_3)\gamma^5 \epsilon\sla^*(\vec{p}_4,\lambda_4)
\frac{1}{2m^0_{N^*}}
\frac{q\sla+m^0_{N^*}}{E-m^0_{N^*}}
p\slas_2 u(\vec{p}_1,\lambda_1) IF_{N^*s}(I).
\end{equation}

\item $N^*(D_{13})$ pole diagram
\be
\nonumber i\kappa\frac{f_{N^*N\eta}f_{N^*N\rho}}{m^2_{\pi}m_{\rho}}
\bar u (\vec{p}_3,\lambda_3) (p\slas_{4}\epsilon^*_\mu(\vec{p}_4,\lambda_4)-p_{4\mu}\epsilon\sla^*(\vec{p}_4,\lambda_4)) \\
\frac{P^{\mu\nu}(q)}{2m_{N^*}(E-m^0_{N^*})}
p_{2_\nu} \gamma^5 p\slas_2
u(\vec{p}_1,\lambda_1) IF_{N^*s}(I)
\ee
$IF_{N^*s}(1/2)=\sqrt{3}.$

\subsection{$\neta \to \dpi$}
\item $N^*(S_{11})$ pole diagram
\begin{equation}
\kappa\frac{f_{N^*N\eta}f_{N^*\Delta\pi}}{m^2_{\pi}}
\bar{u}_{\mu} (\vec{p}_3,\lambda_3) p_4^\mu \gamma^5
\frac{1}{2m^0_{N^*}}
\frac{q\sla+m^0_{N^*}}{E-m^0_{N^*}}
p\slas_2 u(\vec{p}_1,\lambda_1) IF_{N^*s}(I).
\end{equation}

\item $N^*(P_{13})$ pole diagram
\begin{equation}
\kappa \frac{f_{N^*N\eta}f_{N^*\Delta\pi}}{m_\pi^2}
\bar{u}_{\mu} (\vec{p}_3,\lambda_3)\gamma^5 p\slas_4
\frac{1}{2m^0_{N^*}}
\frac{P^{\mu\nu}(q)}{E-m^0_{N^*}}
p_{2\nu} u(\vec{p}_1,\lambda_1) IF_{N^*s}(I).
\end{equation}

\item $N^*(D_{13})$ pole diagram
\begin{equation}
-\kappa \frac{f_{N^*N\eta} f_{N^*\Delta\pi}}{m_\pi^3}
\bar{u}_{\mu} (\vec{p}_3,\lambda_3) p\slas_4
\frac{1}{2m^0_{N^*}}
\frac{P^{\mu\nu}(q)}{E-m^0_{N^*}}
\gamma^5 p\slas_2 p_{2\nu} u(\vec{p}_1,\lambda_1) IF_{N^*s}(I).
\end{equation}
$IF_{N^*s}(1/2)=-\sqrt{2}.$

\subsection{$\nrho \to \nrho$}
\item $N^*(S_{11})$ pole diagram
\begin{equation}
\kappa g_{N^*N\rho}^2
\bar u (\vec{p}_3,\lambda_3)\gamma^5 \epsilon\sla^*(\vec{p}_4,\lambda_4)
\frac{1}{2m^0_{N^*}}
\frac{q\sla+m^0_{N^*}}{E-m^0_{N^*}}
\gamma^5 \epsilon\sla(\vec{p}_2,\lambda_2) u(\vec{p}_1,\lambda_1) IF_{N^*s}(I).
\end{equation}

\item $N^*(D_{13},D_{33})$ pole diagrams
\be
\nonumber \kappa\frac{f_{N^*N\rho}^2}{m_{\rho}^2}
\bar u (\vec{p}_3,\lambda_3) (p\slas_{4}\epsilon^*(\vec{p}_4,\lambda_4)-p_{4\mu}\epsilon\sla^*(\vec{p}_4,\lambda_4)) \\
\times \frac{P^{\mu\nu}(q)}{2m_{N^*}(E-m^0_{N^*})}
(p\slas_{2}\epsilon^*_\nu(\vec{p}_2,\lambda_2)-p_{2\nu}\epsilon\sla^*(\vec{p}_2,\lambda_2))
u(\vec{p}_1,\lambda_1) IF_{N^*s}(I)
\ee
$IF_{N^*s}(1/2)=3,IF_{N^*s}(3/2)=1.$

\subsection{$\nrho \to \dpi$}
\item $N^*(D_{13},D_{33})$ pole diagrams
\be
\nonumber i\kappa\frac{f_{N^*N\rho}f_{N^*\Delta\pi}}{m_{\pi}m_{\rho}}
\bar{u}_{\mu} (\vec{p}_3,\lambda_3) p\slas_4\\
\times \frac{P^{\mu\nu}(q)}{2m_{N^*}(E-m^0_{N^*})}
(p\slas_{2}\epsilon^*_\nu(\vec{p}_2,\lambda_2)-p_{2\nu}\epsilon\sla^*(\vec{p}_2,\lambda_2))
u(\vec{p}_1,\lambda_1) IF_{N^*s}(I)
\ee
$IF_{N^*s}(1/2)=-\sqrt{6},IF_{N^*s}(3/2)=\sqrt{\frac53}.$

\subsection{$\dpi \to \dpi$}
\item $N^*(S_{11},S_{31})$ pole diagrams
\begin{equation}
-\kappa\frac{f_{N^*\Delta\pi}^2}{m^2_{\pi}}
\bar{u}_{\mu} (\vec{p}_3,\lambda_3) p_4^\mu \gamma^5
\frac{1}{2m^0_{N^*}}
\frac{q\sla+m^0_{N^*}}{E-m^0_{N^*}}
\gamma^5 p_2^\nu u_{\nu}(\vec{p}_1,\lambda_1) IF_{N^*s}(I).
\end{equation}

\item $N^*(P_{31})$ pole diagram
\begin{equation}
\kappa\frac{f_{N^*\Delta\pi}^2}{m^2_{\pi}}
\bar{u}_{\mu} (\vec{p}_3,\lambda_3) p_4^\mu
\frac{1}{2m^0_{N^*}}
\frac{q\sla+m^0_{N^*}}{E-m^0_{N^*}}
p_2^\nu u_{\nu}(\vec{p}_1,\lambda_1) IF_{N^*s}(I).
\end{equation}

\item $N^*(P_{13})$ pole diagram
\begin{equation}
\kappa \frac{f_{N^*\Delta\pi}^2}{m_\pi^2}
\bar{u}_{\mu} (\vec{p}_3,\lambda_3)\gamma^5 p\slas_4
\frac{1}{2m^0_{N^*}}
\frac{P^{\mu\nu}(q)}{E-m^0_{N^*}}
\gamma^5 p\slas_2 u_{\nu}(\vec{p}_1,\lambda_1) IF_{N^*s}(I).
\end{equation}

\item $N^*(D_{13},D_{33})$ pole diagrams
\begin{equation}
\kappa \frac{f_{N^*\Delta\pi}^2}{m_\pi^2}
\bar{u}_{\mu} (\vec{p}_3,\lambda_3) p\slas_4
\frac{1}{2m^0_{N^*}}
\frac{P^{\mu\nu}(q)}{E-m^0_{N^*}}
p\slas_2 u_{\nu}(\vec{p}_1,\lambda_1) IF_{N^*s}(I).
\end{equation}
$IF_{N^*s}(1/2)=2,IF_{N^*s}(3/2)=\frac53.$

\end{itemize}

\vfill \eject

\bibliography{refer}

\newpage

\begin{table}
\begin{center}
\begin{tabular}{|l|l|}
\hline
Vertex & ${\cal L}_{int}$ \\
\hline
$NN\pi$
&$-\frac{f_{NN\pi}}{m_{\pi}} \bar{\Psi}\gamma^5\gamma^{\mu} \vec{\tau}\partial_{\mu}\vec{\pi} \Psi $ \\
$N\Delta\pi$
&$\frac{f_{N \Delta\pi}}{m_{\pi}} \bar{\Delta}^{\mu} \vec{S}^{\dagger} \partial_{\mu} \vec{\pi} \Psi + \text{h.c.}$ \\
$\rho\pi\pi$
&$-g_{\rho \pi \pi} (\vec{\pi} \times \partial_{\mu} \vec{\pi}) \vec{\rho}^{\mu} $\\
$NN\rho$
&$-g_{NN\rho} \bar{\Psi}[\gamma^{\mu}-\frac{\kappa_{\rho}}{2m_N}\sigma^{\mu\nu}\partial_{\nu}]\vec{\tau}\vec{\rho}_{\mu} \Psi $\\
$NN\sigma$
&$-g_{NN\sigma} \bar{\Psi} \Psi \sigma $\\
$\sigma\pi\pi$
&$\frac{g_{\sigma \pi \pi}}{2 m_{\pi}} \partial_{\mu} \vec{\pi} \partial^{\mu} \vec{\pi} \sigma $\\
$\sigma\sigma\sigma$
&$-g_{\sigma \sigma \sigma} m_{\sigma} \sigma \sigma \sigma $\\
$NN\rho\pi$
&$\frac{f_{NN\pi}}{m_{\pi}}g_{\rho} \bar{\Psi}\gamma^5\gamma^{\mu}\vec{\tau}\Psi(\vec{\rho}_{\mu} \times \vec{\pi}) $\\
$NNa_1$
&$-\frac{f_{NN\pi}}{m_{\pi}}m_{a_1}\bar{\Psi}\gamma^5\gamma^{\mu}\vec{\tau}\Psi \vec{a}_{\mu} $\\
$a_1\pi\rho$
&$-\frac{g_{\rho}}{m_{a_1}}\left[\partial_{\mu}\vec{\pi}\times\vec{a}_{\nu}-\partial_{\nu}\vec{\pi}\times\vec{a}_{\mu}\right]\left[\partial^{\mu}\vec{\rho}^{\nu}-\partial^{\nu}\vec{\rho}^{\mu}\right] $\\
&$+\frac{g_{\rho}}{2m_{a_1}}\left[\vec{\pi}\times(\partial_{\mu}\vec{\rho}_{\nu}-\partial_{\nu}\vec{\rho}_{\mu})\right]\left[\partial^{\mu}\vec{a}^{\nu}-\partial^{\nu}\vec{a}^{\mu}\right] $\\
$NN\omega$
&$-g_{NN\omega} \bar{\Psi}\gamma^{\mu}\omega_{\mu} \Psi $\\
$\omega\pi\rho$
&$\frac{g_{\omega\pi\rho}}{m_{\omega}}\epsilon_{\alpha\beta\mu\nu}\partial^{\alpha}\vec{\rho}^{ \; \beta}\partial^{\mu}\vec{\pi}\omega^{\nu} $\\
$N \Delta\rho$
&$-i\frac{f_{N\Delta\rho}}{m_{\rho}} \bar{\Delta}^{\mu}\gamma^5\gamma^{\nu}\vec{S}^{\dagger}\vec{\rho}_{\mu\nu}\Psi +\text{h.c.}$\footnotemark[1]\\
$\rho\rho\rho$
&$\frac{g_{\rho}}{2}(\vec{\rho}_{\mu}\times\vec{\rho}_{\nu})\vec{\rho}^{\mu\nu} $\\
$NN \rho\rho$
&$\frac{\kappa_{\rho}g^2_{\rho}}{8m_N}\bar{\Psi}\sigma^{\mu\nu}\vec{\tau}\Psi(\vec{\rho}_{\mu}\times\vec{\rho}_{\nu})$ \\
$\Delta\Delta \pi$
&$\frac{f_{\Delta\Delta\pi}}{m_{\pi}}\bar{\Delta}_{\mu} \gamma^5\gamma^{\nu} \vec{T}\Delta^{\mu}\partial_{\nu}\vec{\pi}$ \\
$\Delta\Delta \rho$
&$-g_{\Delta\Delta\rho}\bar{\Delta}_{\tau} \left(\gamma^{\mu}-i\frac{\kappa_{\Delta\Delta\rho}}{2m_{\Delta}}\sigma^{\mu\nu}\partial_{\nu}\right)\vec{\rho}_{\mu} \vec{T}\Delta^{\tau}$ \\
$NN\eta$
&$-\frac{f_{NN\eta}}{m_{\pi}} \bar{\Psi}\gamma^5\gamma^{\mu} \partial_{\mu}\eta \Psi $ \\
$NNa_0$
&$g_{NNa_0} m_{\pi} \bar{\Psi} \vec{\tau} \Psi \vec{a}_0$ \\
$\pi\eta a_0$
&$g_{\pi\eta a_0} m_{\pi} \eta \vec{\pi} \vec{a}_0$ \\
$N^*(S_{11})N\pi$
&$\frac{f_{N^*N\pi}}{m_{\pi}} \bar{\Psi}_{N^*}\gamma^{\mu}\vec{\tau} \Psi \partial_{\mu}\vec{\pi} + \text{h.c.}$ \\
$N^*(S_{11})N\eta$
&$\frac{f_{N^*N\eta}}{m_{\pi}} \bar{\Psi}_{N^*}\gamma^{\mu} \Psi\partial_{\mu} \eta + \text{h.c.}$ \\
$N^*(S_{11})N\rho$
&$g_{N^*N\rho} \bar{\Psi}_{N^*} \gamma^5 \gamma^{\mu}\vec{\tau}\vec{\rho}_{\mu} \Psi + \text{h.c.}$ \\
$N^*(S_{11})\Delta\pi$
&$\frac{-f_{N^* \Delta\pi}}{m_{\pi}} \bar{\Psi}_{N^*} \gamma^5 \vec{S} {\Delta}^{\mu} \partial_{\mu} \vec{\pi} + \text{h.c.}$ \\
$N^*(P_{13})N\pi$
&$\frac{f_{N^*N\pi}}{m_{\pi}} \bar{\Psi}^\mu_{N^*}\vec{\tau} \Psi \partial_{\mu}\vec{\pi} + \text{h.c.}$ \\
$N^*(P_{13})N\eta$
&$\frac{f_{N^*N\eta}}{m_{\pi}} \bar{\Psi}^\mu_{N^*}\Psi\partial_{\mu} \eta + \text{h.c.}$ \\
$N^*(P_{13})\Delta\pi$
&$\frac{f_{N^* \Delta\pi}}{m_{\pi}} \bar{\Psi}^\mu_{N^*} \gamma^5\gamma^{\nu} \vec{S} {\Delta}_{\mu} \partial_{\nu} \vec{\pi} + \text{h.c.}$ \\
$N^*(D_{13})N\pi$
&$\frac{f_{N^*N\pi}}{m^2_{\pi}} \bar{\Psi} \gamma^5 \gamma^{\nu} \vec{\tau} \Psi_{N^*}^{\mu}\partial_{\nu}\partial_{\mu} \vec{\pi} + \text{h.c.}$ \\
$N^*(D_{13})N\eta$
&$\frac{f_{N^*N\eta}}{m^2_{\pi}} \bar{\Psi} \gamma^5 \gamma^{\nu} \Psi_{N^*}^{\mu}\partial_{\nu}\partial_{\mu} \eta + \text{h.c.}$ \\
$N^*(D_{13})\Delta\pi$
&$i\frac{f_{N^*\Delta\pi}}{m_{\pi}} \bar{\Psi}_{N^*\nu} \vec{S} \gamma^{\mu} \Delta^{\nu} \partial_{\mu} \vec{\pi} +\text{h.c.}$ \\
$N^*(D_{13})N\rho$
&$\frac{f_{N^*N\rho}}{m_{\rho}} \bar{\Psi}_{N^*}^{\mu}\gamma^{\nu}\vec{\tau}\vec{\rho}_{\mu\nu}\Psi +\text{h.c.}$\\
$\Delta^*(S_{31})N\pi$
&$\frac{f_{\Delta^*N\pi}}{m_{\pi}} \bar{\Delta}^*\gamma^{\mu}\vec{S}^\dagger \Psi \partial_{\mu}\vec{\pi} + \text{h.c.}$ \\
$\Delta^*(S_{31})\Delta\pi$
&$\frac{-f_{\Delta^* \Delta\pi}}{m_{\pi}} \bar{\Delta}^* \gamma^5 \vec{T} {\Delta}^{\mu} \partial_{\mu} \vec{\pi} + \text{h.c.}$ \\
$\Delta^*(P_{31})N\pi$
&$-\frac{f_{\Delta^*N\pi}}{m_{\pi}} \bar{\Delta}^* \gamma^5\gamma^{\mu}\vec{S}^\dagger \Psi \partial_{\mu}\vec{\pi} + \text{h.c.}$ \\
$\Delta^*(P_{31})\Delta\pi$
&$-\frac{f_{\Delta^* \Delta\pi}}{m_{\pi}} \bar{\Delta}^*  \vec{T} {\Delta}^{\mu} \partial_{\mu} \vec{\pi} + \text{h.c.}$ \\
$N^*(D_{33})N\pi$
&$\frac{f_{N^*N\pi}}{m^2_{\pi}} \bar{\Psi} \gamma^5 \gamma^{\nu}\vec{S}^\dagger\Psi_{N^*}^{\mu}\partial_{\nu}\partial_{\mu} \vec{\pi} + \text{h.c.}$ \\
$N^*(D_{33})\Delta\pi$
&$i\frac{f_{N^*\Delta\pi}}{m_{\pi}} \bar{\Delta}^*\nu \vec{T} \gamma^{\mu} \Delta^{\nu} \partial_{\mu} \vec{\pi} +\text{h.c.}$ \\
$N^*(D_{33})N\rho$
&$\frac{f_{N^*N\rho}}{m_{\rho}} \bar{\Delta}^{*\mu}\gamma^{\nu}\vec{S}^\dagger\vec{\rho}_{\mu\nu}\Psi +\text{h.c.}$\\
\hline
\end{tabular}
\caption{The effective Lagrangian}\label{tablag}
\end{center}
\end{table}

\begin{table}
\begin{center}
\begin{tabular}{||ll|ll|ll||}
%\hhline{|t:======:t|}
\hline
\multicolumn{4}{||c|}{Mesons}& \multicolumn{2}{c||}{Baryons} \\
\hline
$m_{\pi}$ & 138.03   &$m_{\omega}$ &782.6      &$m_{N}$ & 938.926  \\
$m_{\eta}$ & 547.45   &$m_{a_0}$ & 982.7         &$m_{\Delta}$ & 1232.0   \\
$m_{\sigma}$ & 650.0    &$m_{a_1}$ & 1260.0      &      &          \\
$m_{\rho}$ & 769.0    & &        &        &   \\
%\hhline{|b:======:b|}
\hline
\end{tabular}
\caption{Masses of mesons and baryons (in MeV) used in the calculations.}
\label{masses}
\end{center}
\end{table}

\newpage

\begin{table}
\begin{center}
\label{bg_param}
\begin{tabular}{||l c l c c||}
\hline
%\hhline{|t:=====:t|}
Vertex & Type of the diagram & Coupling constant & Ref. & Cutoff $\Lambda$ [MeV] \\
\hline
\multicolumn{2}{||l}{correlated $\pi\pi$--exchange:} &&&\\
&  {$\rho$--channel} & && {\bf 1000} \\
&  {$\sigma$--channel} & $A_0= $ {\bf 25} $\mathrm {MeV}/F_\pi^2$ && {\bf 900} \\
\hline
$NN\pi$ & {$N$ exchange} & $\frac{f^2_{NN\pi}}{4\pi}=0.0778$ &\cite{Janssen96}& {\bf 1100} \\
%$NN\pi$ & $N$ pole, $m^0_N=1032.33$ & $\frac{f^{(0)\;2}_{NN\pi}}{4\pi}=0.0633$ && {\bf 1200} \\
$N\Delta\pi$ & $\Delta$ exchange & $\frac{f^2_{N\Delta\pi}}{ 4\pi}=0.36$ & \cite{Janssen96}& {\bf 1800} \\
\hline
$NN\rho$ & {$N$ exchange} & $\frac{g^2_{NN\rho}}{ 4\pi} =0.84$ & \cite{Janssen96}& {\bf 1600} \\
& & $\kappa=6.1$ & \cite{Janssen96}&  \\
$NN\rho\pi$ & {contact term} & $\sim f_{NN\pi}g_{NN\rho}$ && {\bf 1100} \\
$NN\pi$ & {$\pi$ exchange} &  $\sim{f_{NN\pi}}$ && {\bf 900} \\
$\pi\pi\rho$ & {$\pi$ exchange} & $\frac{g^2_{\pi\pi\rho}}{ 4\pi}$ && {900} \\
$NN\omega$ & {$\omega$ exchange} & $\frac{g^2_{NN\omega}}{ 4\pi}=11.0$ & \cite{Janssen96}& {\bf 1200} \\
$\omega\pi\rho$ & {$\omega$ exchange} & $\frac{g^2_{\omega\pi\rho}}{ 4\pi}=10.0$ &\cite{Nakayama98,Durso87}& {\bf 1200} \\
$NNa_1$ & $a_1$ exchange & $\sim f_{NN\pi}$&& {\bf 1600} \\
$a_1\pi\rho$ & $a_1$ exchange & $\sim g_{NN\rho}$&& {1600} \\
$NN\rho$ & $\rho$ exchange & $g_{NN\rho},\kappa$ && {\bf 1400} \\
$\rho\rho\rho$ & $\rho$ exchange &  $\sim g_{NN\rho}$ && {1400} \\
$NN\rho\rho$ & contact term & $\sim g^2_{NN\rho}\kappa$ && {\bf 1200} \\
\hline
$N\Delta\pi$ & $N$ exchange & $\frac{f^2_{N\Delta\pi}}{ 4\pi}=0.36$ &\cite{Janssen96}& {\bf 1600} \\
$\Delta\Delta\pi$ & $\Delta$ exchange & $\frac{f^2_{\Delta\Delta\pi}}{ 4\pi}=0.252$ & \cite{Schuetz95J,Brown75}& {\bf 1800} \\
$N\Delta\rho$ & {$\rho$ exchange} & $\frac{f^2_{N\Delta\rho}}{ 4\pi}=20.45$ & \cite{Janssen96}& {\bf 1400} \\
$\Delta\Delta\rho$ & $\rho$ exchange & $\frac{g^{V\,2}_{\Delta\Delta\rho}}{ 4\pi}=4.69$, &\cite{Schuetz95J,Brown75}& {\bf 1400} \\
 & & $\frac{g^T_{\Delta\Delta\rho} }{ g^V_{\Delta\Delta\rho}}=6.1$ & \cite{Schuetz95J,Brown75}& \\
$\pi\pi\rho$ & $\rho$ exchange & $\frac{g^2_{\rho\pi\pi}}{ 4\pi}=2.90$ & \cite{Janssen95}& {\bf 1400} \\ 
\hline
$NN\sigma$ & {$N$ exchange} & $\frac{g^2_{NN\sigma}}{ 4\pi}=13$ &\cite{Durso80}& {\bf 1800} \\
%$NN\pi$ & {$\pi$ exchange} &  $\sim{f_{NN\pi}}$ && {\bf 900} \\
$\pi\pi\sigma$ & { $\pi$ exchange} & $\frac{g^2_{\pi\pi\sigma}}{ 4\pi}=0.25$ & \cite{Krehl97}& {\bf 1050} \\
$NN\sigma$ & {$\sigma$ exchange}& $\sim{g_{NN\sigma}}$ && {\bf 1700} \\
$\sigma\sigma\sigma$ & {$\sigma$ exchange} & $\frac{g^2_{\sigma\sigma\sigma}}{ 4\pi}={\bf 0.275}$ && {1700} \\
\hline
$NN\eta$ & $N$ exchange & $\frac{f^2_{NN\eta}}{ 4\pi}=0.00934$ & \cite{Schuetz98}& {\bf 1500} \\
$NNa_0$ & $a_0$ exchange & $ \frac{g_{NNa_0}g_{\pi\eta a_0}}{ 4\pi}= 8.0$ & \cite{Schuetz98}& {\bf 1500} \\
$\pi\eta a_0$ & \phantom{123}$a_0$ exchange \phantom{123}&  &&\phantom{123}  1500\phantom{123} \\
%\hhline{|b:=====:b|}
\hline
\end{tabular}
\caption{Parameters of the vertices which enter into the background 
diagrams. Free parameters are given in boldface.}
\end{center}
\end{table}

\begin{table}
\begin{center}
\begin{tabular}{||c|c|c|c|c|c||}
%\hhline{~~|t:====:t|}
\cline{3-6}
\multicolumn{2}{l}{} & \multicolumn{4}{|c||}{$f^2/(4\pi)$} \\
%\hhline{|t:==|----||}
\hline
Resonance & Bare mass [MeV]&$\pi N$& $\pi\Delta$ &$\rho  N$ &$\eta N$ \\
\hline
$N^*_{S_{11}}(1535) $ &2051 & 0.00045& 1.09(-)& & 0.0247\\
\hline
$N^*_{S_{11}}(1650) $ &1919 & 0.0067& & 0.046\footnotemark&\\
\hline
$N^*_{P_{13}}(1720) $ &1910 & 0.0031& 0.0085(-)& &0.079(-)\\
\hline
$N^*_{D_{13}}(1520) $ &2263 & 0.00037& 0.0118&0.609 &0.0008 \\
\hline
$\Delta(1232)$ & 1459 & 0.163& & &\\
\hline
$\Delta_{S_{31}}(1620)$ & 2419 &0.0154 & 2.91(-)& &\\
\hline
$\Delta_{P_{31}}(1910)$ & 2121 & 0.0043& 0.007(-)& &\\
\hline
$\Delta_{D_{33}}(1700)$ & 2252 & 0.00038& 0.03(-)& 0.011&\\
%\hhline{|b:======:b|}
\hline
\end{tabular}
\caption{Parameters of the  pole graphs: bare masses and coupling
constants. The minus
sign in parenthesis indicates that the coupling
constant is negative.}
\label{pole_param}
\end{center}
\end{table}

\begin{table}[h]
\begin{center}
\begin{tabular}{||c c c c||}
%\hhline{|t:====:t|}
\hline
\vrule height 15pt depth 5pt width 0pt
 & this work   &  Ref. \cite{Koch80}& SM95 \cite{SM95}\\
\hline
\vrule height 15pt depth 5pt width 0pt
$S_{11}$ &  {0.195}  &  0.173 $\pm$ 0.003 & 0.175 \\
\vrule height 15pt depth 5pt width 0pt
$S_{31}$ & {--0.110} & --0.101 $\pm$ 0.004 & --0.087 \\
\vrule height 15pt depth 5pt width 0pt
$P_{11}$ & {--0.089} & --0.081 $\pm$ 0.002 & --0.068 \\
\vrule height 15pt depth 5pt width 0pt
$P_{31}$ & {--0.046} & --0.045 $\pm$ 0.002 & --0.039 \\
\vrule height 15pt depth 5pt width 0pt
$P_{13}$ & {--0.031} & --0.030 $\pm$ 0.002 & --0.022 \\
\vrule height 15pt depth 5pt width 0pt
$P_{33}$ &  {0.209}  & 0.214 $\pm$ 0.002 & 0.209 \\
%\hhline{|b:====:b|}
\hline
\end{tabular}
\caption{ The $s$ and $p$-wave $\pi N$ scattering lengths
and volumes
in terms of $m_{\pi^+}^{-(2L+1)}$.}  
\label{scatlength}
\end{center}
\end{table}

\end{document}